\documentclass[12pt]{iopart}
\usepackage{cite}

\usepackage[utf8]{inputenc}
\usepackage[T1]{fontenc}
\usepackage[english]{babel}

\usepackage{graphicx}
\usepackage{subfigure}

\usepackage{amsmath}
\usepackage{amssymb}

\usepackage{bm}

\usepackage{units}
\usepackage[squaren]{SIunits}

\usepackage[version=3]{mhchem}

\usepackage{acronym}

\usepackage{hyperref}

\newcommand{\ee}{\ensuremath{\mathrm{e}}} 
\newcommand{\ii}{\ensuremath{\mathrm{i}}} 

\newcommand{\abs}[1]{\ensuremath{\left\vert #1 \right\vert}} 
 
\newcommand{\eto}[1]{\ensuremath{\,{\ee}^{#1}}} 
\renewcommand{\vec}[1]{{\ensuremath{\bm{#1}}}} 
\newcommand{\dd}{\ensuremath{\mathrm{d}}} 
 
\newcommand{\firstpderiv}[2]{\ensuremath{\frac{\partial #1}{\partial #2}}} 

\newcommand{\qmprod}[2]{\ensuremath{\left\langle #1\middle\vert #2 \right\rangle}} 

\DeclareMathOperator{\Real}{Re} 
\DeclareMathOperator{\Imag}{Im} 
\DeclareMathOperator{\sign}{sign} 

\begin{document}

\title[A variational approach to Bogoliubov excitations and dynamics
  of dipolar BECs]{A variational approach to Bogoliubov excitations and
  dynamics of dipolar Bose-Einstein condensates}
\author{Manuel Kreibich, J\"org Main and G\"unter Wunner}
\address{1. Institut f\"ur Theoretische Physik, Universit\"at
  Stuttgart, 70550 Stuttgart, Germany}

\begin{abstract}
  We investigate the stability properties and the dynamics of
  Bose-Einstein condensates with axial symmetry, especially with
  dipolar long-range interaction, using both simulations on grids and
  variational calculations.  We present an extended variational ansatz
  which is applicable for axial symmetry and show that this ansatz can
  reproduce the lowest eigenfrequencies of the Bogoliubov spectrum,
  and also the corresponding eigenfunctions.  Our variational ansatz
  is capable of describing the roton instability of pancake-shaped
  dipolar condensates for arbitrary angular momenta.  After
  investigating the linear regime we apply the ansatz to determine the
  dynamics and show how the angular collapse is correctly described
  within the variational framework.
\end{abstract}

\pacs{03.75.Kk, 67.85.De, 05.45.-a}


\acrodef{BDGE}{Bogoliubov-de Gennes equations}
\acrodef{BEC}{Bose-Einstein condensate}
\acrodef{DDI}{dipole-dipole interaction}
\acrodef{GPE}{Gross-Pitaevskii equation}
\acrodef{ITE}{imaginary time evolution}
\acrodef{LRI}{long-range interaction}
\acrodef{SRI}{short-range interaction}
\acrodef{TDVP}{time-dependent variational principle}

\section{Introduction}
\label{sec:introduction}

As an alternative to the direct numerical solution of the \ac{GPE} and
the \ac{BDGE}, which describe the ground states and excitations of
\acp{BEC} \cite{Pitaevskii03a}, Rau \etal \cite{Rau10a,Rau10b}
proposed a variational approach with coupled Gaussians to calculate
stationary solutions and the stability of a condensate using the
methods of nonlinear dynamics. Specifically, they used the \ac{TDVP}
to map the \ac{GPE} to a nonlinear dynamical system for the
variational parameters whose fixed points correspond to ground states
of the \ac{GPE}. The stability can be analysed by means of the
linearised equations of motion, i.\,e.\ in terms of the eigenvalues of
the Jacobian. However, it remained unclear whether there is a direct
relationship between the eigenvalues of the \ac{BDGE} and those of the
Jacobian. In a recent work \cite{Kreibich12a} we analysed this problem
for spherically symmetric systems and showed that a variational ansatz
with coupled Gaussians has limitations in that it can only describe
excitations with an angular momentum $l \leq 2$. We present an
extended variational ansatz and show that there is indeed a good
agreement between the lowest eigenvalues of the \ac{BDGE} and the
Jacobian for arbitrary angular momenta, with or without additional
\ac{LRI}.

A variational ansatz with a single Gaussian \cite{Perez-Garcia97a}
cannot reproduce the biconcave structure of dipolar condensates
revealed in certain parameter regions \cite{Ronen07a,Dutta07a}. Using
the variational ansatz of coupled Gaussians, however, it was possible
to obtain non-Gaussians shapes \cite{Rau10a,Rau10b,Rau10c} and to
reproduce the wave function and the energy of structured ground
states. Additionally, the stability was analysed which revealed that
the collapse breaks the axial symmetry within the variational
approach. However, the ansatz used in Ref.~\cite{Rau10c} is only
capable of describing modes with angular momentum $m=0$ and $m=2$ (see
discussion in \cite{Kreibich12a}). The angular collapse with $m=3$
symmetry, calculated numerically on a grid in \cite{Wilson09a}, is
therefore not accessible to this variational ansatz of coupled
Gaussians.

Several extensions of the Gaussian ansatz have been proposed in the
literature \cite{Buccoliero07a, Buccoliero09a, Maucher10a}, but they
allow for no systematic inclusion of arbitrary angular momenta. It is
the purpose of this work to extend the variational ansatz of coupled
Gaussians in such a way that an instability with an arbitrary
projection of the angular momentum on the $z$ axis can be described
for systems with axial symmetry. We present an extended variational
ansatz including angular exponentials $\exp(\ii m \phi)$, which are
eigenfunctions of the $z$ component of the angular momentum
operator. This enables us to describe modes with arbitrary angular
momentum within the variational framework, which will be demonstrated
by applying the ansatz to condensates with contact interaction only,
and with \ac{DDI}.

In Sec.~\ref{sec:stab-excit} we discuss the stability and excitations
of \acp{BEC}. We first present the method of numerically solving the
\ac{BDGE}, which we need for a comparison with the variational
method. The extended variational ansatz is described and applied to
calculate the eigenfrequencies of elementary excitations. We compare
the results with those obtained from the full-numerical solution. In
Sec.~\ref{sec:dynamics} we apply the variational ansatz to calculate
the time evolution of a dipolar \ac{BEC} and demonstrate that the
ansatz can describe the angular collapse. In
Sec.~\ref{sec:conclusion-outlook} we draw conclusions and give an
outlook on future work.

\section{Stability and excitations}
\label{sec:stab-excit}

The dime-dependence of a \ac{BEC} is described in the mean-field
approximation by the time-dependent \ac{GPE} \cite{Pitaevskii03a}
\begin{align}
  \label{eq:tgpe}
  \ii \hbar \firstpderiv{\psi}{t}(\vec{r},t) = \left[ -
    \frac{\hbar^2}{2M} \Delta + \frac{M}{2} \omega_\rho^2 \left(
      \rho^2 + \lambda^2 z^2 \right) + N \Phi_\text{int}(\vec{r},t)
  \right] \psi(\vec{r},t),
\end{align}
where $N$ is the particle number,
\begin{align}
  \label{eq:potscat}
  \Phi_\text{int}(\vec{r},t) = \int \dd^3 r^\prime \,
  V_\text{int}(\vec{r}-\vec{r}^\prime) \abs{\psi(\vec{r}^\prime,t)}^2
\end{align}
the mean field produced by the inter-particle interaction,
$\psi(\vec{r},t)$ the wave function of the condensate normalised to
unity, and $M$ the mass of the atom species. The harmonic external
trap is given by the trapping frequencies $\omega_\rho$ and $\omega_z$
in the $\rho$ and $z$ direction, and can be characterised by the trap
aspect ratio $\lambda = \omega_z / \omega_\rho$. We measure all
lengths in units of the harmonic oscillator length $a_\text{ho} \equiv
\sqrt{\hbar / M \omega_\rho}$, all frequencies in units of
$\omega_\rho$, and the time in units of $1/\omega_\rho$. A stationary
solution of Eq.~(\ref{eq:tgpe}) has the time dependence
$\psi(\vec{r},t) = \psi(\vec{r}) \exp(-\ii \mu t / \hbar)$, where
$\mu$ is the chemical potential of the condensate.

The interaction potential is given by \cite{Stuhler05a}
\begin{align}
  \label{eq:1}
  V_\text{int}(\vec{r}-\vec{r}^\prime) = \frac{4 \pi a \hbar^2}{M}
  \delta(\vec{r}-\vec{r}^\prime) + \frac{\mu_0 \mu_\text{m}^2}{4 \pi}
  \frac{1-3\frac{(z-z^\prime)^2}{(\vec{r}-\vec{r}^\prime)^2}}
  {\abs{\vec{r}-\vec{r}^\prime}^3}.
\end{align}
The quantity $a$ is the s-wave scattering length and $\mu_\text{m}$
the magnetic moment of the atom species and is, e.\,g., $\mu_\text{m}
= 6 \mu_\text{B}$ ($\mu_\text{B}$ the Bohr magneton) for \ce{^{52}Cr}
\cite{Griesmaier05a}. The dipoles are assumed to be aligned along the
$z$ axis by an external magnetic field. The interactions can be
characterised by the dimensionless parameters $N a / a_\text{ho}$ and
$D \equiv N \frac{\mu_0}{4 \pi} \frac{M}{\hbar^2}
\frac{\mu_\text{m}^2}{a_\text{ho}}$ \cite{Ronen06a}. In this article,
we investigate \acp{BEC} with and without \ac{LRI}. For the condensate
with \ac{DDI} we use values of $D=30$ and $\lambda=7$, since for these
parameters structured ground states appear \cite{Ronen07a,Rau10b}. For
the \ac{BEC} without \ac{LRI} we use a trap aspect ratio of $\lambda =
\sqrt{8}$, which was used in experiments \cite{Jin96a}. The scattering
length can be tuned in a wide range via Feshbach resonances
\cite{Inouye98a}. Thus, we use this quantity as a free parameter in
our calculations.

\subsection{Full-numerical treatment}
\label{sec:full-numer-treatm}

Besides the variational approach we also perform grid calculations to
directly solve the \ac{GPE} and \ac{BDGE}, which allows us to compare
both methods. Before solving the \ac{BDGE}, one first needs to find
the ground state of a \ac{BEC}\@. In this work, we use the \ac{ITE}
for this task. We evolve an initial wave function on a grid in
imaginary time $\tau$ defined by $\tau \equiv \ii t$, and as time
evolves the wave function converges to the ground state. The time
evolution on the grid is done via the split operator method
\cite{Feit82a}, in which the time evolution operator in imaginary time
is given by
\begin{align}
  \eto{-\hat{H} \Delta\tau / \hbar} = \eto{-\frac{1}{2} \hat{T}
    \Delta\tau / \hbar} \eto{-\hat{V} \Delta\tau / \hbar}
  \eto{-\frac{1}{2} \hat{T} \Delta\tau / \hbar} + {\cal
    O}(\Delta\tau^3),
\end{align}
where $\hat{T}$ and $\hat{V}$ represent the kinetic and the potential
terms of the mean-field Hamiltonian, respectively. The action of the
kinetic part is trivial in Fourier space, whereas the action of the
potential part is trivial in real space. Since we are dealing with a
cylindrically symmetric system, we can use the Fourier-Hankel
algorithm of \cite{Ronen06a} to compute the necessary Fourier
transforms for a time step and the dipolar integrals via the
convolution theorem.

To derive the \ac{BDGE}, one starts from the usual ansatz for the
perturbation of a stationary state $\psi_0$ with chemical potential
$\mu$ \cite{Pitaevskii03a}
\begin{align}
  \label{eq:bdgansatz}
  \psi(\vec{r},t) = \left[ \psi_0(\vec{r}) + \lambda \left( u(\vec{r})
      \eto{-\ii \omega t} + v^*(\vec{r}) \eto{\ii \omega^* t} \right)
  \right] \eto{-\ii \mu t / \hbar},
\end{align}
where $\omega$ is the frequency and $\lambda$ the amplitude of the
perturbation ($\abs{\lambda} \ll 1$). After inserting this ansatz into
the time-dependent \ac{GPE}~(\ref{eq:tgpe}), while neglecting terms of
second order in $\lambda$, and collecting terms evolving in time with
$\exp(-\ii \omega t)$ and $\exp(\ii \omega t)$, respectively, one
obtains the \ac{BDGE}
\begin{subequations}
  \label{eq:bdg}
  \begin{align}
    \hbar \omega u(\vec{r}) = &\left[ H_0 - \mu + \int \dd^3 r^\prime
      \, V_\text{int}(\vec{r}-\vec{r}^\prime)
      \abs{\psi_0(\vec{r}^\prime)}^2 \right]
    u(\vec{r}) \nonumber \\
    &+ \int \dd^3 r^\prime \, V_\text{int}(\vec{r}-\vec{r}^\prime)
    \psi_0^*(\vec{r}^\prime) u(\vec{r}^\prime) \psi_0(\vec{r})
    \nonumber \\
    &+ \int \dd^3 r^\prime \, V_\text{int}(\vec{r}-\vec{r}^\prime)
    \psi_0(\vec{r}^\prime) v(\vec{r}^\prime) \psi_0(\vec{r}),
    \displaybreak[0] \\
    -\hbar \omega v(\vec{r}) = &\left[ H_0 - \mu + \int \dd^3 r^\prime
      \, V_\text{int}(\vec{r}-\vec{r}^\prime)
      \abs{\psi_0(\vec{r}^\prime)}^2 \right]
    v(\vec{r}) \nonumber \\
    &+ \int \dd^3 r^\prime \, V_\text{int}(\vec{r}-\vec{r}^\prime)
    \psi_0^*(\vec{r}^\prime) u(\vec{r}^\prime) \psi_0^*(\vec{r})
    \nonumber \\
    &+ \int \dd^3 r^\prime \, V_\text{int}(\vec{r}-\vec{r}^\prime)
    \psi_0(\vec{r}^\prime) v(\vec{r}^\prime) \psi_0^*(\vec{r}),
  \end{align}
\end{subequations}
where $H_0 = -\frac{\hbar^2}{2M} \Delta + \frac{M}{2} \omega_\rho^2
(\rho^2 + \lambda^2 z^2)$. Due to the cylindrical symmetry, we can
make an ansatz for the solutions
\begin{align}
  \label{eq:bdgsep}
  u(\vec{r}) = \eto{\ii m \phi} u(\rho,z), &&
  v(\vec{r}) = \eto{\ii m \phi} v(\rho,z),
\end{align}
where $m$ is the usual quantum number of the projection of the angular
momentum operator on the $z$ axis. Since the Laplacian in cylindrical
coordinates applied to the ansatz~(\ref{eq:bdgsep}) yields
\begin{align}
  \Delta \eto{\ii m \phi} u(\rho,z) =
  \left(
    \Delta_{\rho,z}
    - \frac{m^2}{\rho^2}
  \right) \eto{\ii m \phi} u(\rho,z), 
\end{align}
which only depends on $\abs{m}$, there is a degeneracy of the
eigenmodes for $\abs{m}>0$. Thus, we can restrict ourselves to
solutions with $m \geq 0$. Another symmetry of the system is the
reflection at the $z=0$ plane ($z \to -z$). Therefore, the solutions
can be described by the parity $\pi_z = \pm 1$ under this
reflection. We denote the set of quantum numbers with
$m^{\pi_z}$. There are always solutions of the \ac{BDGE} for
$\omega=0$ and the trapping frequencies, which represent the gauge
mode and the centre of mass oscillations
\cite{Pitaevskii03a,Kreibich12a}.

We solve the linear, nonlocal \ac{BDGE}~(\ref{eq:bdg}) by calculating
the matrix elements of the right-hand side in coordinate
representation by means of the Fourier-Hankel algorithm
\cite{Ronen06a}. The lowest eigenvalues of the resulting
high-dimensional matrix are then calculated using the Arnoldi
iteration \cite{Arnoldi51a}, implemented in the Fortran ARPACK
routines \cite{Lehoucq97a}.

The solutions of the \ac{BDGE}~(\ref{eq:bdg}) yield the frequencies
and the shape of elementary oscillations of the condensates. But it
may happen that the Bogoliubov spectrum of the ground state of a
dipolar \ac{BEC} contains imaginary frequencies \cite{Wilson09a},
which correspond to dynamical instabilities. A small perturbation of
the ground state then leads to an exponential decay.

\begin{figure}[t]
\centering
  \includegraphics[width=0.7\columnwidth]{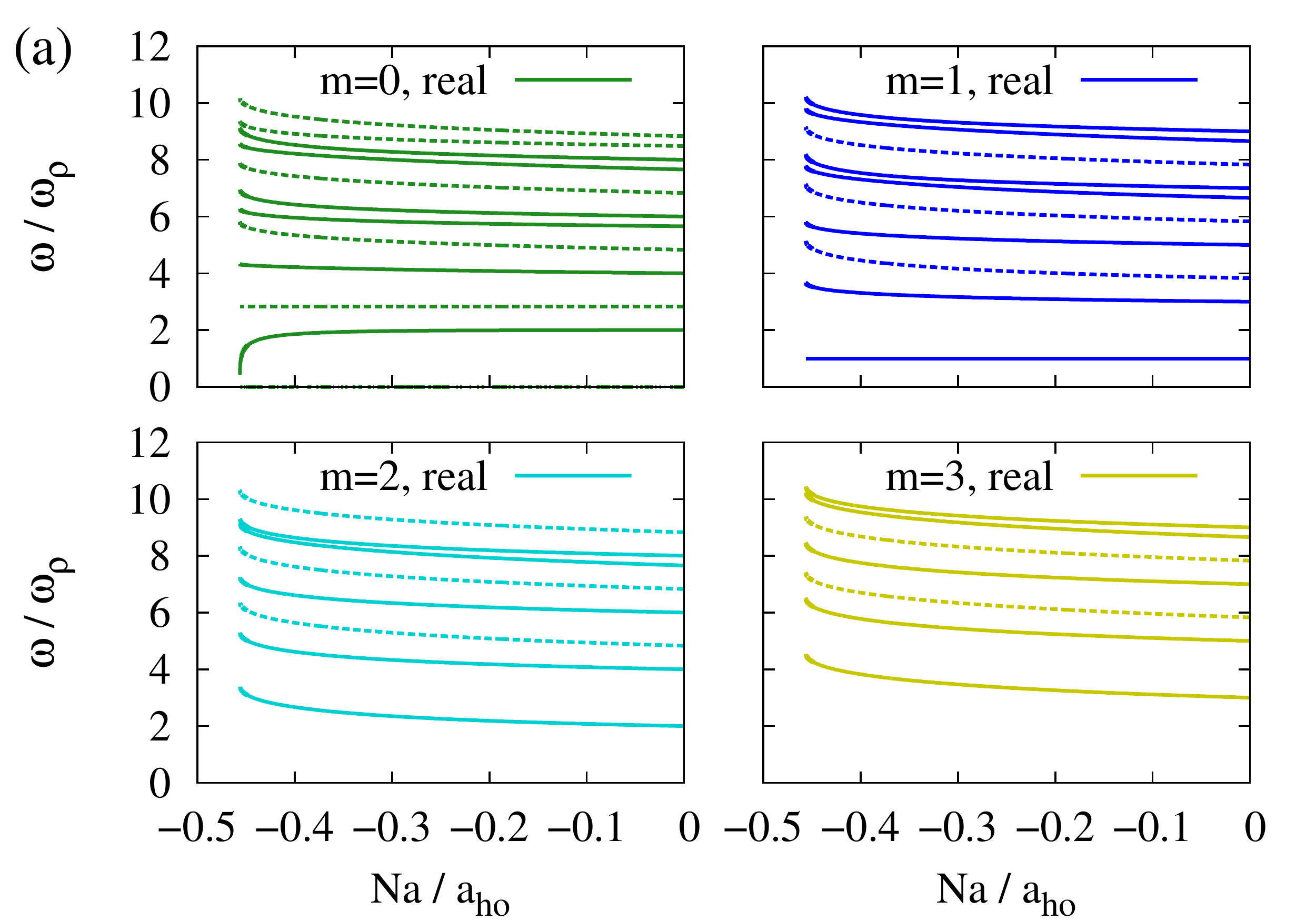}
  \includegraphics[width=0.7\columnwidth]{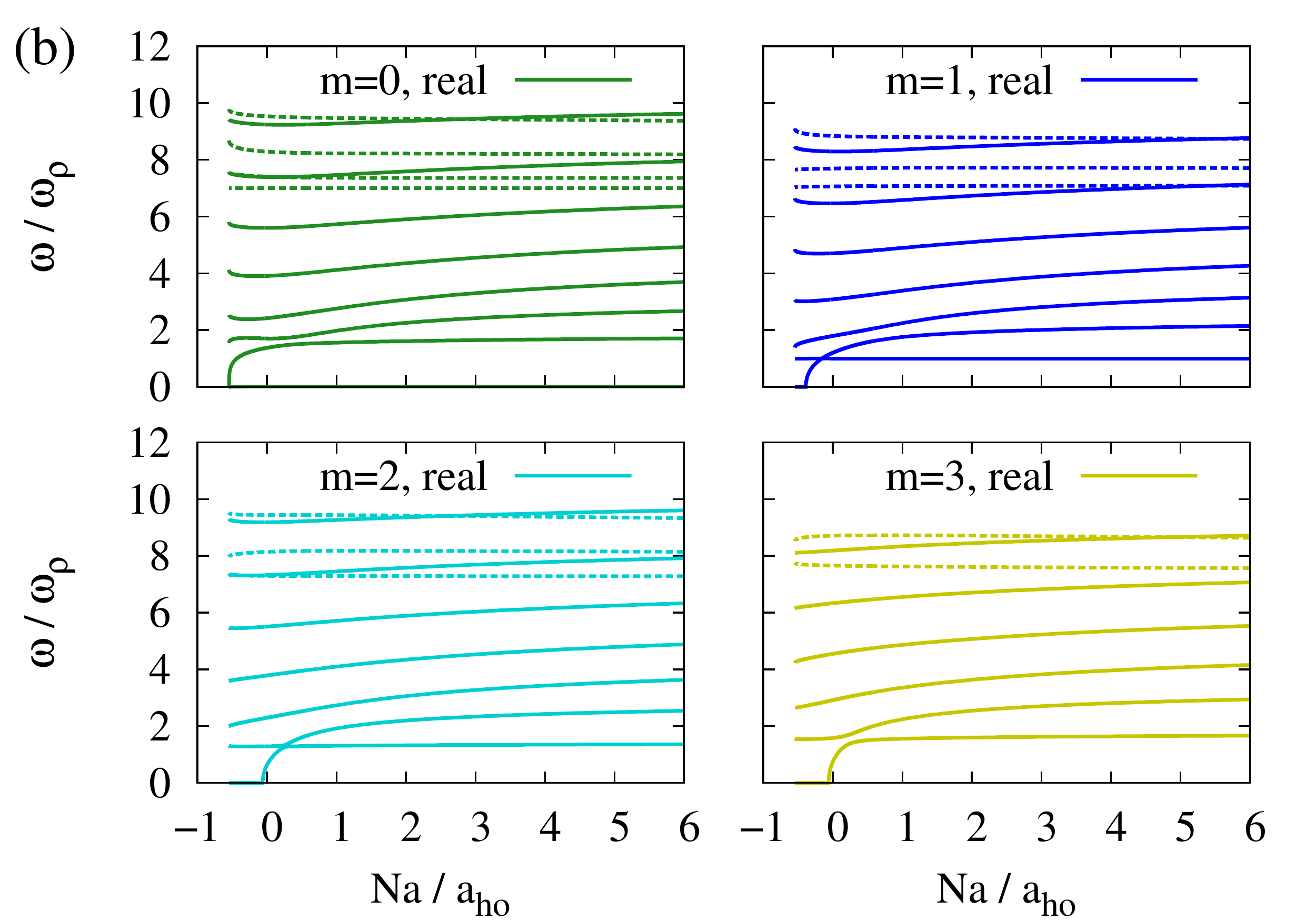}
  \caption{Bogoliubov spectra of the ground states for (a) a \ac{BEC}
    with short-range interaction ($\lambda=\sqrt{8}, D=0$) and (b) a
    dipolar \ac{BEC} ($\lambda=7, D=30$) as a function of the
    scattering length $Na / a_\text{ho}$ (shown are the real parts of
    the frequencies) with even (solid) and odd (dashed) modes. Both
    spectra feature the lowest $m=0$ mode, whose frequency goes to
    zero when the scattering length reaches the critical value
    $a_\text{crit}$, and the centre of mass oscillations at the
    trapping frequencies. Unique for the dipolar \ac{BEC} are the
    lowest excitations for higher angular momenta $m > 0$, which turn
    unstable and mark a local collapse of the condensate.}
  \label{fig:pancake_bogo}
\end{figure}

We calculated the ground states and Bogoliubov spectra for a \ac{BEC}
with an attractive \ac{SRI} only ($D=0$) and a trap aspect ration of
$\lambda = \sqrt{8}$, and for a dipolar \ac{BEC} with $\lambda = 7$
and $D=30$, for which it was shown that the structured ground state
changes its stability in a pitchfork bifurcation \cite{Rau10c} at the
scattering length $Na/a_\text{ho} \approx -0.215$\footnote{In
  \cite{Rau10c} a Gaussian ansatz with $N_\text{G} = 5$ Gaussians was
  used. The scattering length, where the bifurcation takes place,
  however, converges slowly with increasing number of Gaussians and
  may be different in our calculations.}. The structured ground
states, where the maximum of particle density lies away from the
centre, appear in isolated regions in parameter space with $\lambda =
7, 11, 15, \dots$ \cite{Ronen07a}. For our choice of parameters,
$\lambda=7$ and $D=30$, the ground state assumes a biconcave structure
for the scattering lengths $-0.546 \lesssim Na/a_\text{ho} \lesssim
1.4$.

For both systems ground states only exist for scattering lengths
larger than the critical scattering length, $a \geq
a_\text{crit}$. For the condensate with \ac{SRI}, the critical
scattering length is given by $N a_\text{crit} / a_\text{ho} \approx
-0.457$. \acp{BEC} at small negative scattering lengths may exist,
since the kinetic energy compensates the attractive interaction and
stabilises the condensate. The critical scattering length was probed
experimentally in a condensate of \ce{^{85}Rb}, and significant
deviations from the results of the \ac{GPE} were measured
\cite{Roberts01a}, which may be explained by taking into account
higher-order nonlinear effects \cite{Gammal01a}. The critical
scattering length of a dipolar \ac{BEC} considerably depends on the
trap geometry, in our case ($\lambda=7$, $D=30$) its value is given by
$N a_\text{crit} / a_\text{ho} \approx -0.546$. A good agreement
between experiment and the results of the \ac{GPE} was found
\cite{Koch08a}.

Fig.~\ref{fig:pancake_bogo} shows the Bogoliubov spectra. Since both
systems have already been considered in the literature (see, e.\,g.,
\cite{Edwards96a} for the \ac{BEC} with \ac{SRI}, and \cite{Ronen06a}
for the dipolar \ac{BEC}), we only give a short review. The spectrum
of the \ac{BEC} without \ac{LRI} (Fig.~\ref{fig:pancake_bogo}a) can be
understood by taking the limit $a \to 0$, which yields the exact
eigenvalues of the cylindrically symmetric harmonic oscillator
\begin{align}
  \frac{\omega_{n_\rho,n_z,m}}{\omega_\rho} = 2 n_\rho + \abs{m} +
  \lambda n_z,
\end{align}
where $n_{\rho,z} \in \mathbb{N}_0$ are the quantum numbers for
excitations along the $\rho$ and $z$ direction, respectively. When
decreasing the scattering length towards the critical value, below
which no stationary solutions exist anymore, the frequencies vary
slightly, which means that in this regime the dynamics is dominated by
the external trap, and the interaction acts as a quasi
perturbation. Just slightly above the critical point the lowest mode
with $m=0$ drops to zero, which marks the global collapse of the
condensate. The oscillations of the centre of mass, which constantly
lie at the trapping frequencies, are represented by the lowest odd
$m=0$ mode ($\omega / \omega_\rho = \sqrt{8}$) and the lowest even
$m=1$ mode ($\omega / \omega_\rho = 1$).

\begin{figure}[t]
\centering
  \includegraphics[width=0.7\columnwidth]{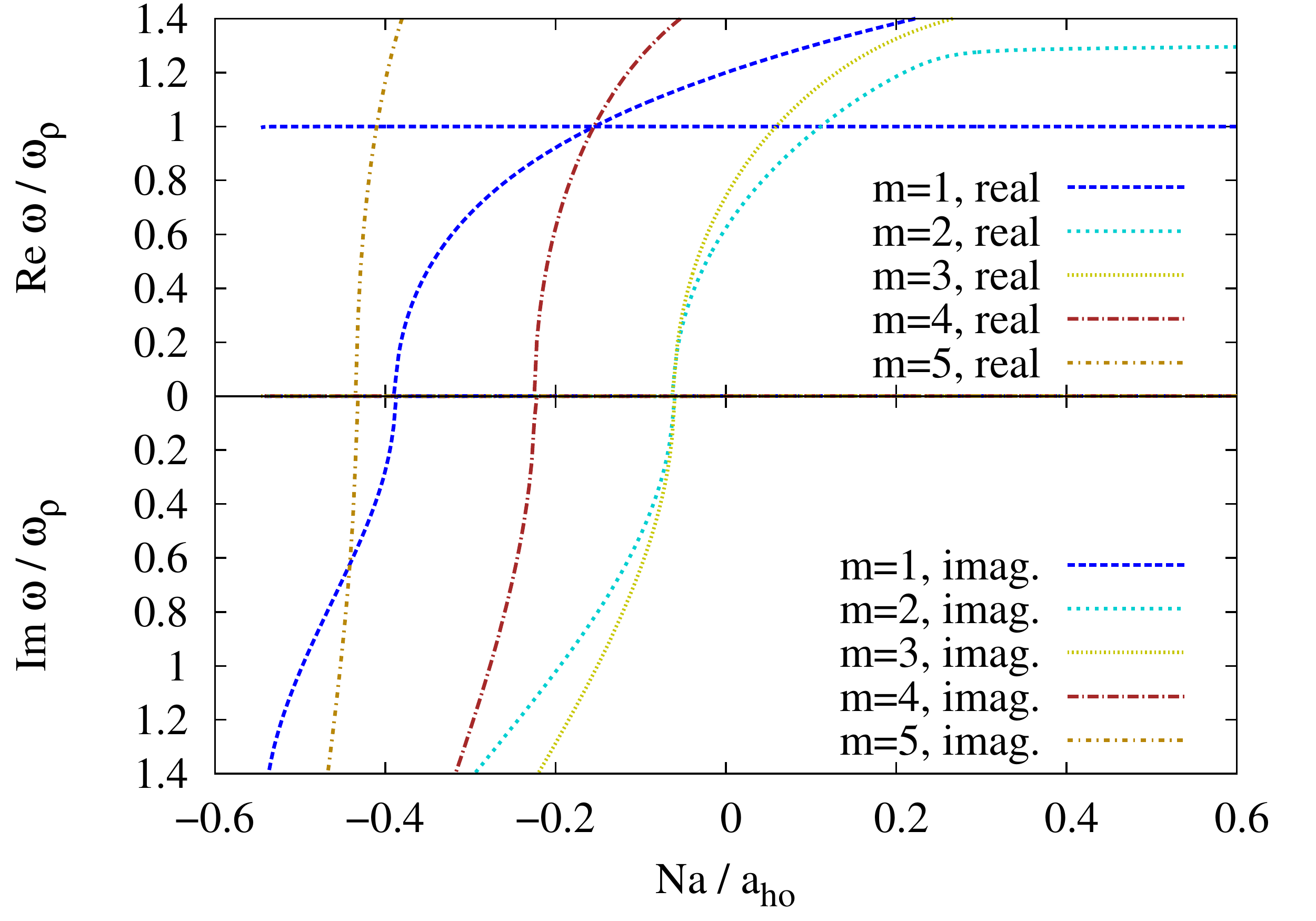}
  \caption{The lowest modes of the Bogoliubov spectrum of the dipolar
    \ac{BEC} from Fig.~\ref{fig:pancake_bogo}b. Shown are the real and
    also imaginary parts of the frequencies. In our resolution of the
    scattering length, the modes with $m=2$ and $m=3$ turn unstable
    simultaneously, but the latter has a higher imaginary frequency
    below the critical scattering length, which means that this mode
    dominates the collapse in the vicinity of the stationary
    solution.}
  \label{fig:dipolar_bogo_crit}
\end{figure}

The Bogoliubov spectrum of the dipolar condensate
(Fig.~\ref{fig:pancake_bogo}b) indicates a stable condensate when the
scattering length is far from the critical point. However, for
decreasing scattering length, not only the mode with $m=0$ but also
frequencies for higher angular momenta decrease and drop to zero. At
these points, those modes become unstable, as the eigenfrequencies are
imaginary. Fig.~\ref{fig:dipolar_bogo_crit} shows the real and
imaginary parts of the lowest eigenfrequencies for the angular momenta
$m=1,\dots,5$. The modes with $m=2$ and $m=3$ are the first ones which
turn unstable, and thus mark the local collapse of the condensate.
This means, that collapse occurs via local density-fluctuations in
contrast to a global collapse, where the condensate collapse to the
trap centre \cite{Wilson09a}. In Sec.~\ref{sec:dynamics} we discuss
the dynamics of such a local collapse, with the result that an
extended variational ansatz is necessary to describe the
collapse. This is a unique feature of dipolar \acp{BEC} and is related
to the biconcave structure of the ground state occurring for specific
trapping geometries \cite{Rau10c,Wilson09a}. These instabilities are
referred to as rotons, or, more specifically, as discrete ``angular
rotons'', since these excitations assume a shape with azimuthal nodal
surfaces \cite{Ronen07a}.

Worth mentioning also is the avoided crossing in the lowest modes for
the angular momenta $m=0,2,3$, whereas for $m=1$ there is a regular
crossing since $\omega = \omega_\rho$ is an exact solution for every
scattering length, which does not allow for an avoided crossing.

\subsection{Variational approach}
\label{sec:variational-approach}

The grid calculations are very accurate, if grid size and resolution
are chosen carefully, but computationally expensive. The variational
method provides an alternative in that the computation time reduces
significantly compared to grid calculations. With this motivation, Rau
\etal \cite{Rau10a,Rau10b,Rau10c} used the variational ansatz
\begin{align}
  \label{eq:varold}
  \psi = \sum\limits_{k=1}^{N_\text{G}} \eto{\ii(A_x^k x^2 +A_y^k y^2
    +A_z^k z^2 +\gamma^k)}
\end{align}
to calculate stationary solutions and excitations of dipolar
condensates. Here, $N_\text{G}$ is the number of Gaussians, $A_x^k$,
$A_y^k$ and $A_z^k$ describe the widths and $\gamma^k$ the amplitude
of each Gaussian. The quantities $(A_x^k, A_y^k, A_z^k, \gamma^k)$ are
the variational parameters of this ansatz. In a recent article
\cite{Kreibich12a} we showed that the variational ansatz with coupled
Gaussians can only describe modes with an angular momentum of $l \leq
2$, $\abs{m} \leq 2$. For that reason we proposed an extended
variational ansatz with coupled Gaussians and spherical harmonics,
which can describe modes with arbitrary angular momenta in spherically
symmetric systems. We found a good agreement of the lowest eigenmodes
with those obtained by grid calculations.

For the systems with axial symmetry considered here, this ansatz is
not applicable, since the angular momentum is no longer a good quantum
number and therefore the angular part of the excitations is not
described by spherical harmonics. However, the $z$ component of the
angular momentum, represented by the quantum number $m$ and the
eigenfunctions $\exp(\ii m \phi)$, is still conserved. Using this
knowledge we construct an extended variational ansatz, which is
specific to cylindrically symmetric systems, and which is given by
\begin{align}
  \label{eq:varansatz}
  \psi = \sum\limits_{k=1}^{N_\text{G}} \left( 1 + \sum\limits_{m\neq
      0} \sum\limits_{p=0,1} d_{m,p}^k \rho^{\abs{m}} z^p \eto{\ii m
      \phi} \right) \eto{-A_\rho^k \rho^2 -A_z^k z^2 -p_z^k z
    -\gamma^k}.
\end{align}
The complex variational parameters $A_\rho^k$ and $A_z^k$ describe the
width in the $\rho$ and $z$ direction, $p_z^k$ is the displacement
along the $z$ axis, and $\gamma^k$ describes the amplitude of each
Gaussian. With these parameters alone, the wave function does not
depend on the angular coordinate $\phi$, thus the parameters
$(A_\rho^k, A_z^k, p_z^k, \gamma^k)$ are responsible only for any
$m=0$ excitation. To account for higher angular momenta and arbitrary
parity $\pi_z$, we include the sums over $m$ and $p$ in the brackets
($\pi_z=+1$ belongs to $p=0$, and $\pi_z=-1$ to $p=1$). The amplitude
of each of these excitations is given by the variational parameters
$d_{m,p}^k$.

To derive the equations of motion for the variational parameters, we
follow the procedure in \cite{Rau10a,Kreibich12a} and apply the
Dirac-Frankel-McLachlan \ac{TDVP} \cite{McLachlan64a,Dirac30a}. It
states that the norm of the difference between the left- and the
right-hand side of the Schr\"odinger or \ac{GPE}
\begin{align}
  I = || \ii \chi(t) - \hat{H} \psi(t) ||^2
\end{align}
must be minimised for a given variational ansatz $\psi =
\psi(\vec{z})$ with respect to the variational parameters
$\vec{z}$. The variation of $\chi$, while $\psi$ is fixed, and the
identification of $\chi$ with $\dot{\psi}$ leads to the necessary
condition \cite{Cartarius08b}
\begin{align}
  \label{eq:tdvpeom}
  \mathbf{K} \dot{\vec{z}} = -\ii \vec{h},
\end{align}
where the matrix $\mathbf{K}$ and the vector $\vec{h}$ are defined by
\begin{align}
  \label{eq:tdvpkh}
  K_{ij} = \qmprod{\firstpderiv{\psi}{z_i}}{\firstpderiv{\psi}{z_j}},
  && h_i = \qmprod{\firstpderiv{\psi}{z_i}}{\hat{H} \psi}.
\end{align}
The remaining task is to evaluate all integrals appearing in
Eq.~(\ref{eq:tdvpkh}). Technical details are presented
in~\ref{sec:calc-integr}. Since the ground state in the \ac{GPE}
corresponds to a fixed point of the equations of motion
(\ref{eq:tdvpeom}), we can find the ground state of a system by a
nonlinear root search, for which we require
\begin{align}
  \dot{z}_i =
  \begin{cases}
    \ii \mu & \text{for } z_i \equiv \gamma^k, \\
    0 & \text{else},
  \end{cases}
\end{align}
where $\mu$ is the chemical potential.

Stability and excitations in the framework of the \ac{TDVP} are given
by the linearised equations of motion, for which one first needs to
rewrite the complex equations of motion~(\ref{eq:tdvpeom}) into real
ones by defining the new vector of variational parameters
$\tilde{\vec{z}}$ as $\tilde{\vec{z}} = (\Real \vec{z}, \Imag
\vec{z})$. The time-dependence of a small perturbation
$\delta\tilde{\vec{z}}$ is then given by \cite{Rau10a}
\begin{align}
  \delta\dot{\tilde{z}}_i = \sum\limits_j \left.
    \firstpderiv{\dot{\tilde{z}}_i}{\tilde{z}_j}
  \right|_{\tilde{\vec{z}}=\tilde{\vec{z}}_0} \delta\tilde{z}_j \equiv
  \sum\limits_j J_{ij} \delta\tilde{z}_j,
\end{align}
with the Jacobian $\mathbf{J}$ evaluated at the fixed point
$\tilde{\vec{z}}_0$. These linearised equations of motion are as usual
solved by a harmonic time-dependence of the perturbation
\begin{align}
  \delta\tilde{\vec{z}}(t) = \eto{\ii \omega t}
  \delta\tilde{\vec{z}}(0),
\end{align}
which leads to the eigenvalue problem
\begin{align}
  \ii \omega \delta\tilde{\vec{z}}(0) = \mathbf{J}
  \delta\tilde{\vec{z}}(0).
\end{align}
A real $\omega$ belongs to an oscillatory motion, as for real
eigenvalues $\omega$ of the \ac{BDGE}. The symmetry of the system,
which leads to the quantum numbers $m$ and $\pi_z$ in the Bogoliubov
spectrum, gives rise to a block structure of the Jacobian which is
further explained in~\ref{sec:structure-jacobian}.

The eigenvalues of the Jacobian and the \ac{BDGE} can be directly
compared. But is there also a relationship between the eigenvectors of
the Jacobian $\delta\tilde{\vec{z}}$ and the eigenfunctions of the
\ac{BDGE} $u$ and $v$? For a given eigenvector
$\delta\tilde{\vec{z}}$, a wave function can be constructed by taking
the difference of the excitation with variational parameters
$\tilde{\vec{z}}_0 + \delta\tilde{\vec{z}}$ and the wave function of
the stationary solution with the parameters $\tilde{\vec{z}}_0$,
\begin{align}
  \label{eq:varwavefunc}
  \delta\psi(\tilde{\vec{z}}_0,\delta\tilde{\vec{z}}) \equiv
  \psi(\tilde{\vec{z}}_0 + \delta\tilde{\vec{z}}) -
  \psi(\tilde{\vec{z}}_0).
\end{align}
Since we are dealing with an excitation in the linear regime, one can
Taylor expand the first summand in Eq.~(\ref{eq:varwavefunc}) to first
order in $\delta\tilde{\vec{z}}$. This yields the function
\cite{Rau10a}
\begin{align}
  \delta\psi = \delta\tilde{\vec{z}} \cdot \left.
    \firstpderiv{\psi}{\tilde{\vec{z}}}
  \right|_{\tilde{\vec{z}}=\tilde{\vec{z}}_0},
\end{align}
which is the scalar product of the eigenvector and the gradient of the
wave function with respect to the variational parameters evaluated at
the stationary solution. Thus, every eigenvector
$\delta\tilde{\vec{z}}$ belongs uniquely to a function $\delta\psi$.

As the Jacobian $\mathbf{J}$ is real and a stable mode is represented
by an imaginary eigenvalue $\ii\omega$, for every eigenvalue $\ii
\omega$ with $\omega > 0$ there is also an eigenvalue $-\ii \omega$,
therefore we can construct corresponding functions $\delta\psi_+$ and
$\delta\psi_-$ which evolve in time as $\exp(\ii \omega t)$ and
$\exp(-\ii \omega t)$, respectively. They can be combined to the
perturbation
\begin{align}
  \delta\psi = \delta\psi_- \eto{-\ii \omega t} + \delta\psi_+
  \eto{\ii \omega t}.
\end{align}
Comparing this with the Bogoliubov ansatz from
Eq.~(\ref{eq:bdgansatz}) leads to the relationship
\begin{align}
  u \leftrightarrow \delta\psi_-, && v^* \leftrightarrow \delta\psi_+,
\end{align}
which links the eigenvectors of the Jacobian with the eigenfunctions
of the \ac{BDGE}.

\subsection{Results for the \ac{BEC} with scattering interaction}
\label{sec:results-bec-with}

\begin{figure}[t]
\centering
  \includegraphics[width=0.7\columnwidth]{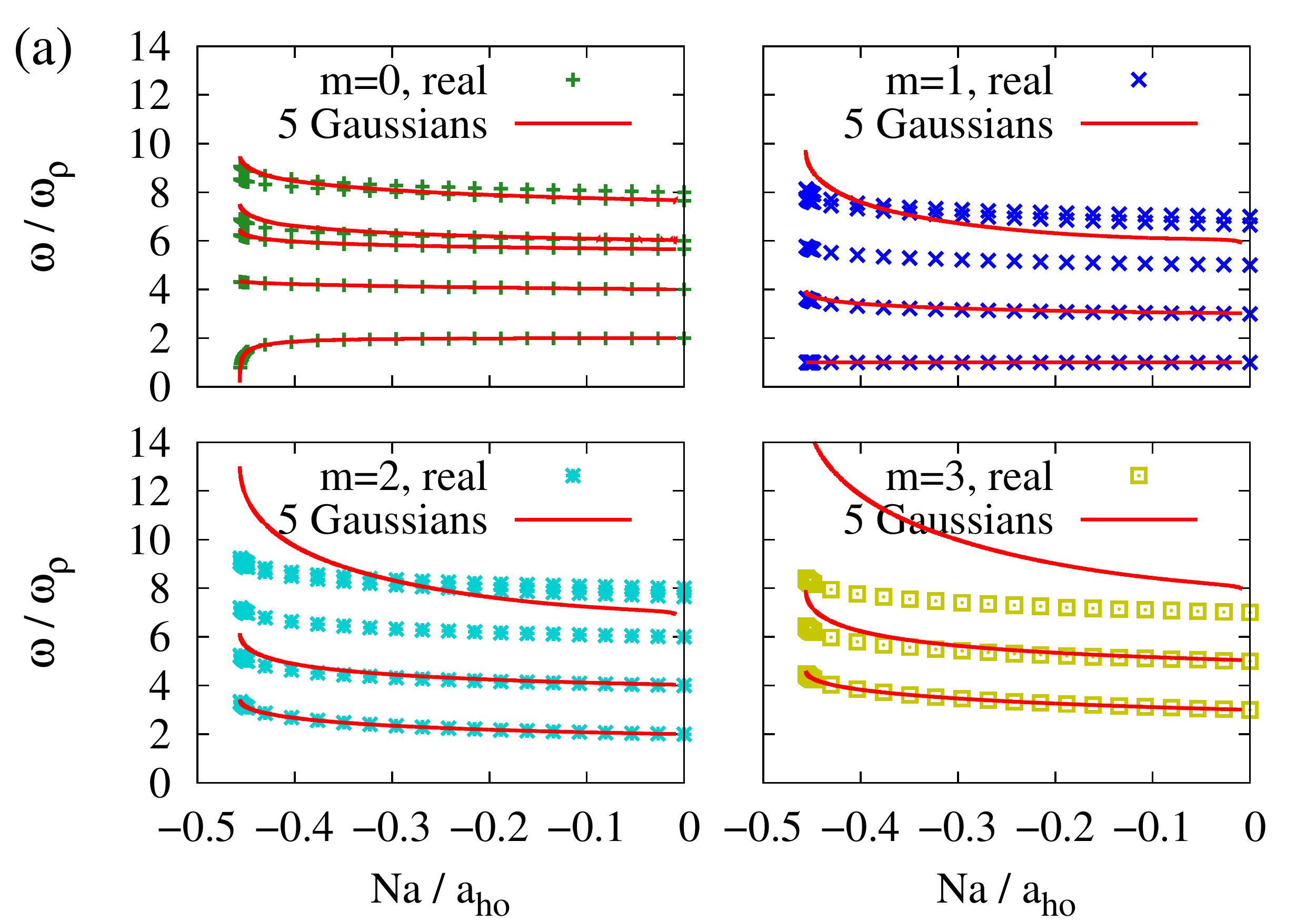}
  \includegraphics[width=0.7\columnwidth]{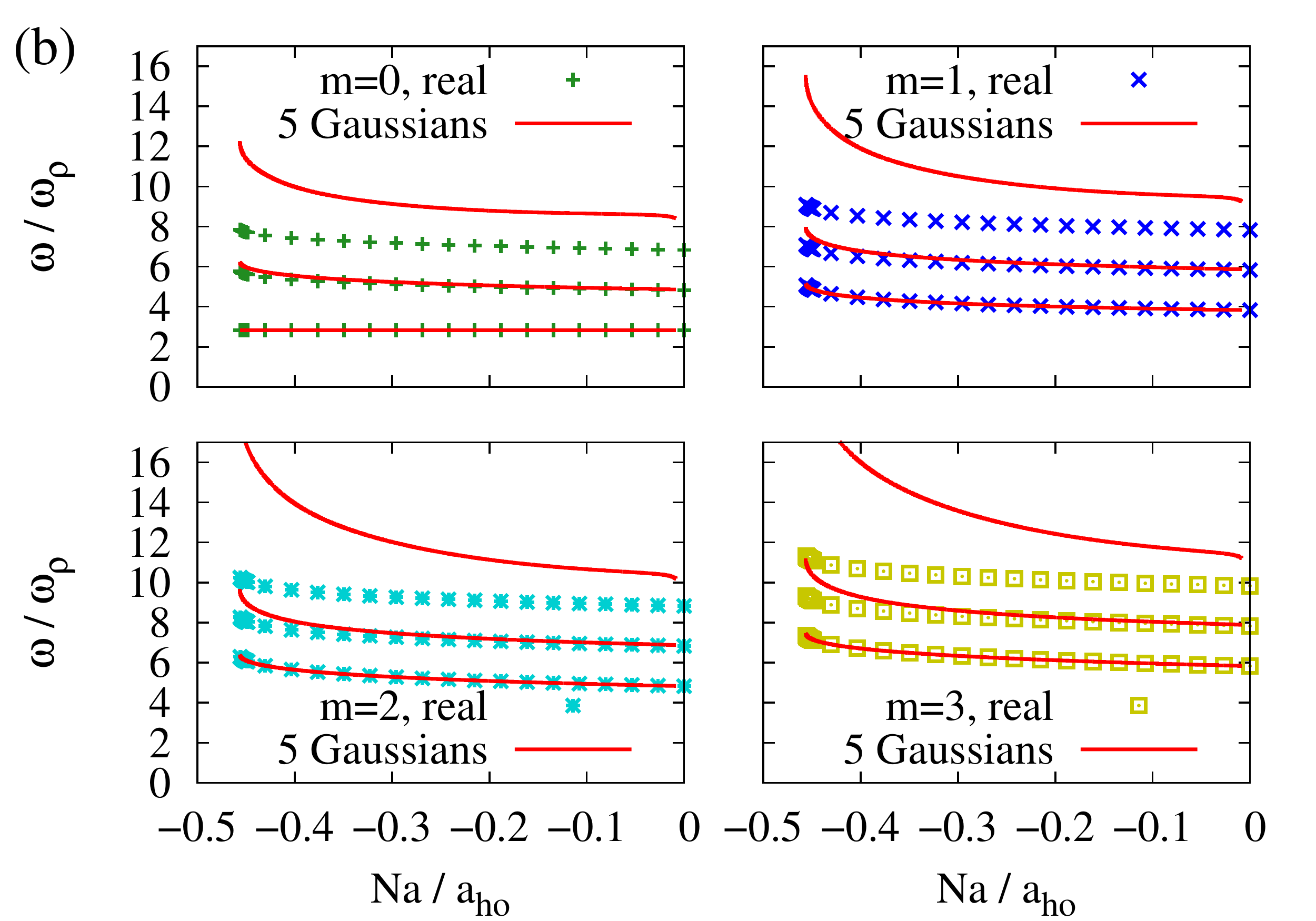}
  \caption{Comparison of the Bogoliubov spectrum for a \ac{BEC} with
    \ac{SRI} from Fig.~\ref{fig:pancake_bogo}a with the eigenvalues of
    the Jacobian obtained from the variational ansatz for even (a) and
    odd (b) parity. We find a very good agreement for the lowest two
    modes of each angular momentum and parity, only close to the
    critical scattering length deviations of the second lowest modes
    for $m>0$ can be seen, which become larger for increasing angular
    momentum.}
  \label{fig:pancake_bogo_5gauss}
\end{figure}

We now apply the variational ansatz~(\ref{eq:varansatz}) to a
condensate with an attractive \ac{SRI}.
Fig.~\ref{fig:pancake_bogo_5gauss} shows the results, where
$N_\text{G}=5$ coupled Gaussians have been used (solid lines),
compared to the full-numerical calculations (dotted lines). For the
sake of clarity we distinguish between even modes in
Fig.~\ref{fig:pancake_bogo_5gauss}a and odd modes in
Fig.~\ref{fig:pancake_bogo_5gauss}b. One recognises that the lowest
mode of each angular momentum and parity can perfectly be reproduced
by the variational ansatz. Interestingly, as with our previous ansatz
for spherically symmetric systems \cite{Kreibich12a}, the frequencies
of the centre of mass oscillations (lowest odd $m=0$ at $\omega =
\sqrt{8} \omega_\rho$, lowest even $m=1$ at $\omega = \omega_\rho$)
agree, within numerical accuracy, with an arbitrary number of coupled
Gaussians, even with $N_\text{G}=1$. For the second lowest modes we
also find a very good agreement as long as the scattering length is
not close to the critical value. For the case $m=0$, $\pi_z=1$ even
more eigenvalues of the Jacobian are close to the numerical results.

\begin{figure}
\centering
  \includegraphics[width=0.7\columnwidth]{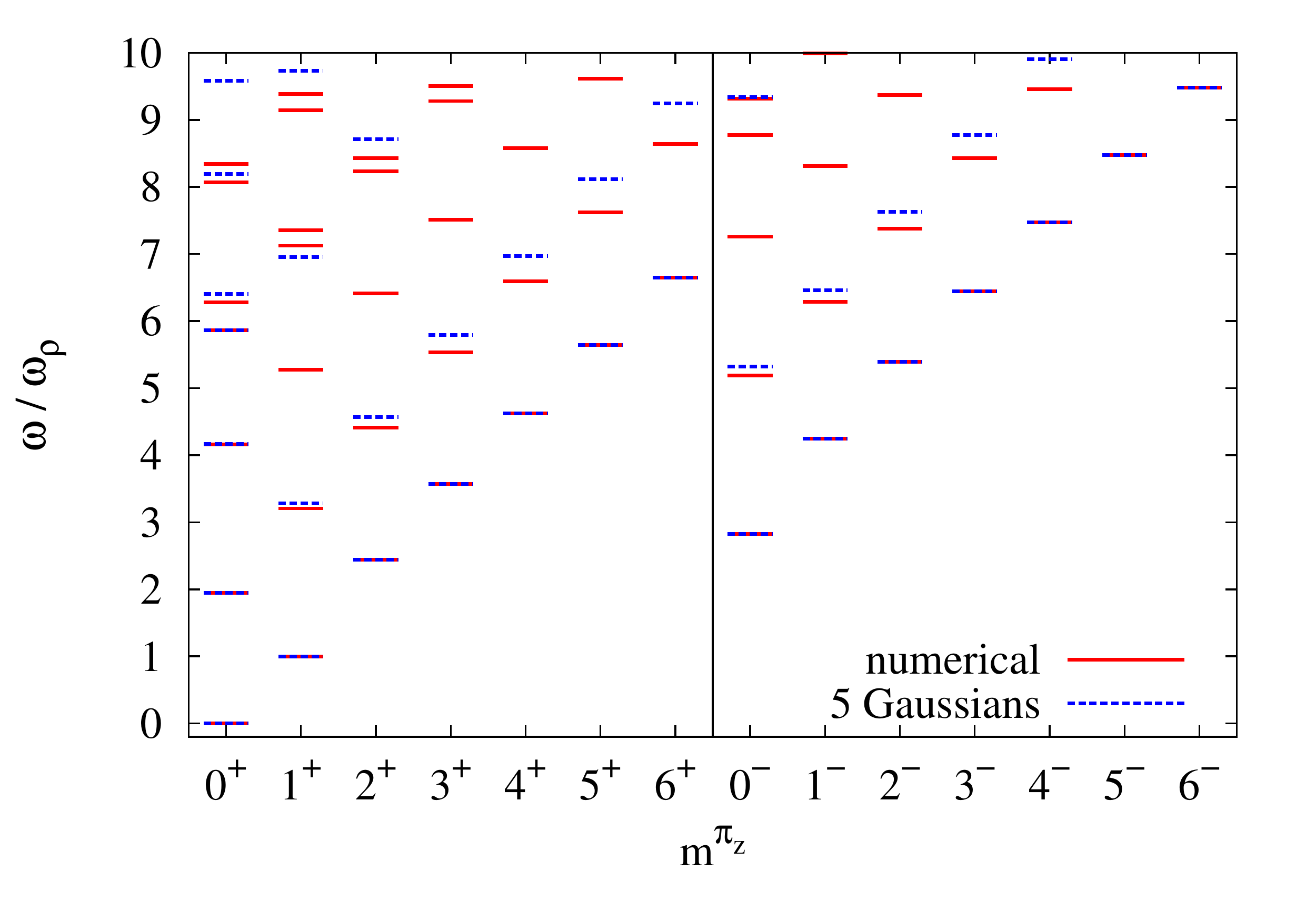}
  \caption{Comparison of both spectra as in
    Fig.~\ref{fig:pancake_bogo_5gauss}, but here for a fixed
    scattering length of $Na/a_\text{ho}=-0.34$ and angular momenta up
    to $m=6$. The variational ansatz also works for these modes and
    can describe the lowest modes of the Bogoliubov spectrum,
    therefore we are able to establish a connection between both
    methods.}
  \label{fig:pancake_spectrum}
\end{figure}

To demonstrate the power of our extended ansatz, we calculated
eigenmodes for angular momenta up to $m=6$ (see
Fig.~\ref{fig:pancake_spectrum}) for a fixed scattering length
$Na/a_\text{ho}=-0.34$. Even for these angular momenta the lowest
modes agree well. For the second excitations and lower angular momenta
we find also a good agreement, however, the deviations of the
variational and numerical results become larger for increasing angular
momenta. Nevertheless, there is an obvious relationship between the
eigenfrequencies of the Bogoliubov spectrum and the eigenvalues of the
Jacobian, where our extended variational ansatz~(\ref{eq:varansatz})
-- specific to cylindrically symmetric systems -- has been used. Thus,
the variational ansatz is a valid alternative to simulations on grids
if one is interested only in the lowest modes.

\subsection{Results for the dipolar \ac{BEC}}
\label{sec:results-dipolar-bec}

\begin{figure}[t]
\centering
  \includegraphics[width=0.7\columnwidth]{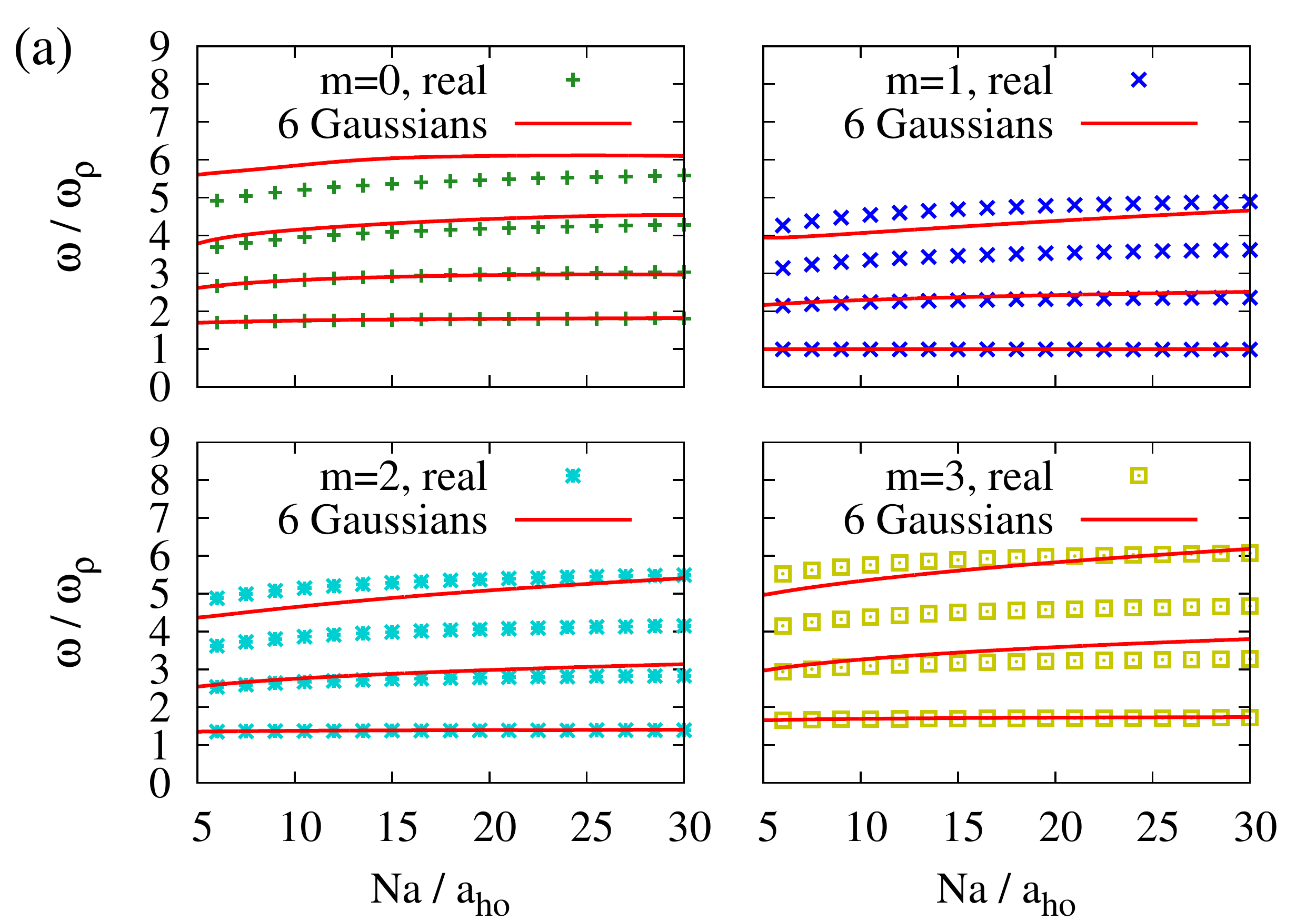}
  \includegraphics[width=0.7\columnwidth]{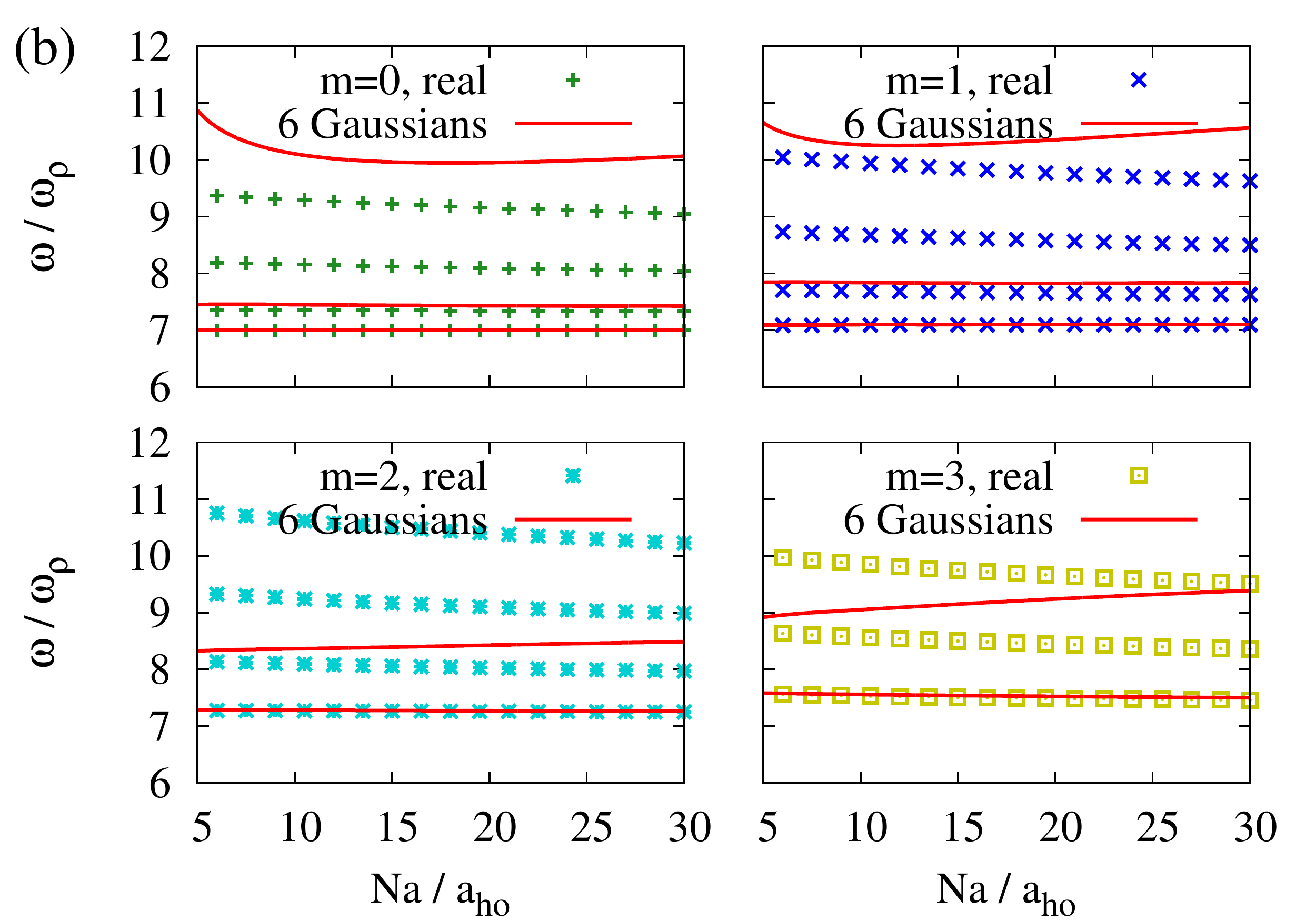}
  \caption{Comparison of the Bogoliubov spectrum for a dipolar
    \ac{BEC} from Fig.~\ref{fig:pancake_bogo}b with the eigenvalues of
    the Jacobian obtained from the variational ansatz. Shown are the
    modes with even (a) and odd (b) parity in the stable regime with a
    scattering length $Na/a_\text{ho} \geq 5$. Whereas the two lowest
    modes agree well for even parity, for odd parity the second lowest
    modes differ quite clearly, especially for higher angular momenta
    $m \geq 2$.}
  \label{fig:dipolar_bogo_6gauss}
\end{figure}

We now want to include the \ac{DDI} in the calculations and especially
verify if the extended variational ansatz can describe the stability
change for higher angular momenta $m>0$. We first investigate the
eigenmodes in the stable regime with a scattering length
$Na/a_\text{ho} \geq 5$. Fig.~\ref{fig:dipolar_bogo_6gauss} shows the
comparison of both spectra for the angular momenta $m=0,\dots,3$. With
$6$ coupled Gaussians, the lowest modes of each angular momentum and
parity (solid lines) are converged to their corresponding numerical
frequencies (dotted lines). As in the case of the \ac{BEC} without
\ac{LRI}, the centre of mass oscillations agree within numerical
accuracy. For the modes with even parity, there is a good agreement
even for the second lowest modes, where the differences become larger
for rising scattering length and angular momentum. In the case of odd
parity, the deviations for the second lowest modes become quite large
for $m \geq 2$, when the scattering length is large.

\begin{figure}[t]
\centering
  \includegraphics[width=0.7\columnwidth]{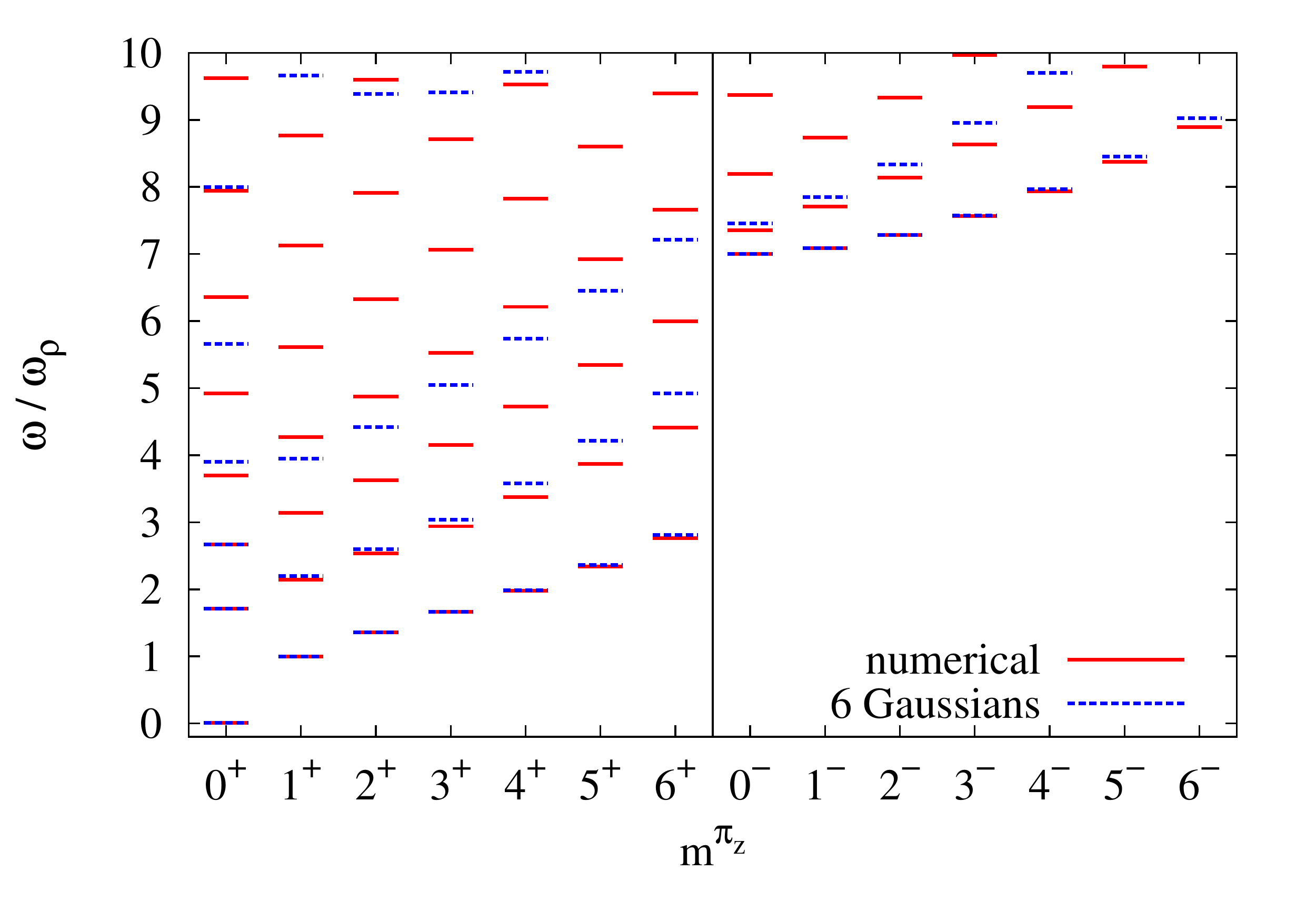}
  \caption{Comparison of both spectra as in
    Fig.~\ref{fig:dipolar_bogo_6gauss}, but here for a fixed
    scattering length of $Na/a_\text{ho} = 6$. For the lowest mode of
    each angular momentum and parity, there is a very good
    agreement. For the second lowest modes we find differences, which
    get larger for increasing angular momenta. The correspondence is
    slightly better for even than for odd modes. As in the case
    without \ac{LRI} there is a link between the full-numerical
    Bogoliubov spectrum and the eigenvalues of the Jacobian resulting
    from the variational ansatz.}
  \label{fig:dipolar_spectrum}
\end{figure}

We have also investigated eigenmodes with an angular momentum up to
$m=6$ in the stable regime. Fig.~\ref{fig:dipolar_spectrum} shows the
results for a fixed scattering length $Na/a_\text{ho} = 6$. The
variational ansatz can reproduce the lowest modes even for these
angular momenta and, since the scattering length is not too large,
also the second lowest modes. For the case $m=0$, $\pi_z=1$, we find
the lowest three modes converged, and a good agreement of the fourth
lowest mode. These investigations show that the variational
ansatz~(\ref{eq:varansatz}) can also reproduce the full-numerical
Bogoliubov spectrum, if only the lowest modes are compared.

\begin{figure}[t]
\centering
  \includegraphics[width=0.7\columnwidth]{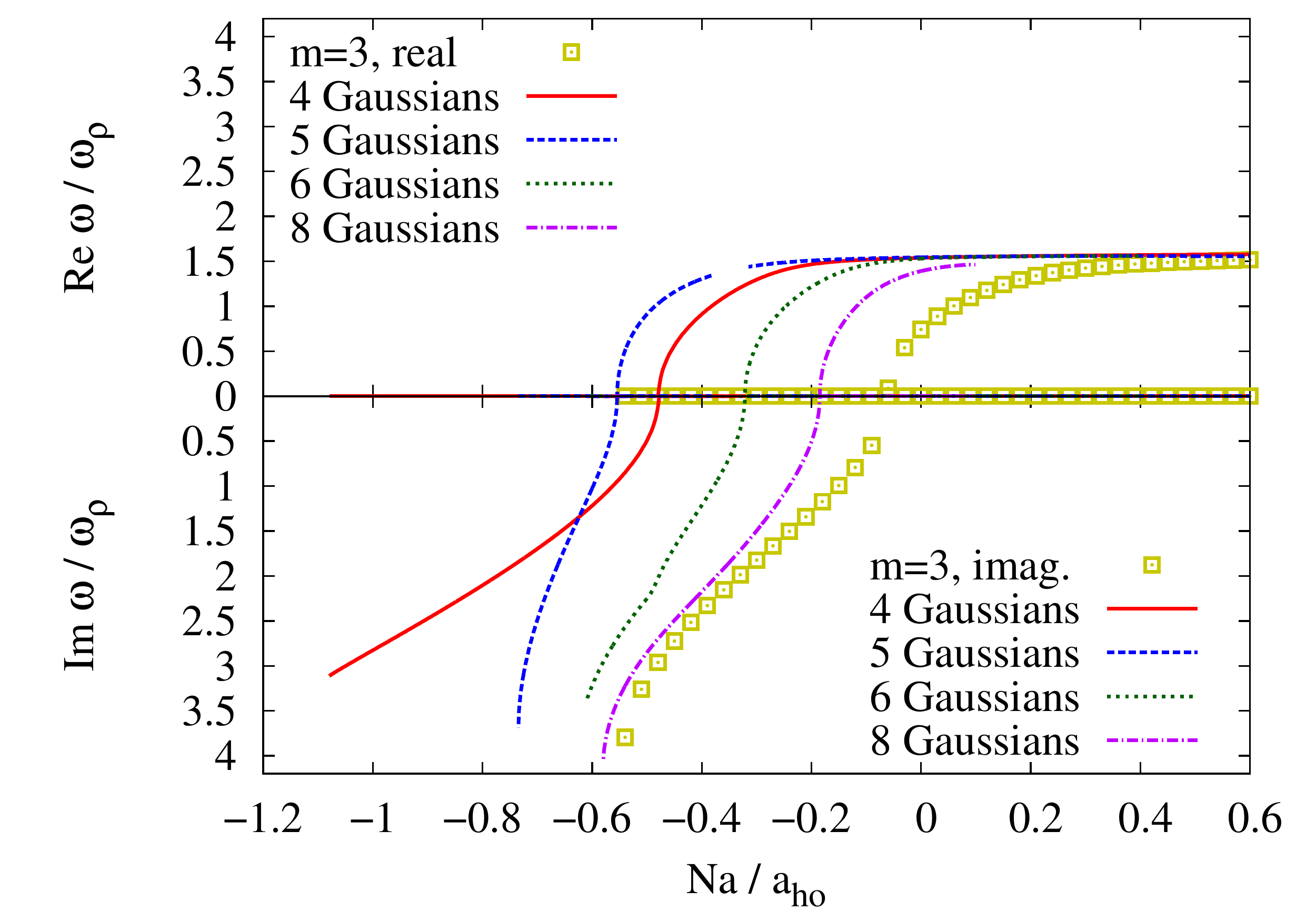}
  \caption{Lowest mode with $m=3$ which turns unstable when the
    scattering length is decreased. Compared are the full-numerical
    frequencies (dotted line) with the results from the variational
    ansatz with different number of Gaussians (solid and dashed
    lines). The accuracy of the variational calculations increases
    gradually with the number of coupled Gaussians.}
  \label{fig:dipolar_bogo_gauss_crit}
\end{figure}

We now investigate the interesting case of a small negative scattering
length, where the Bogoliubov spectrum shows unstable roton modes for
different angular momenta (see Fig.~\ref{fig:dipolar_bogo_crit}). We
focus here on the case $m=3$, which cannot be described by the ansatz
of coupled Gaussians Eq.~(\ref{eq:varold}), hence the extended
variational ansatz is
necessary. Fig.~\ref{fig:dipolar_bogo_gauss_crit} shows the lowest
modes with $m=3$ obtained using different numbers of Gaussians (solid
and dashed lines). For comparison the full-numerical results are also
plotted (dotted line). We find that with an increasing number of
coupled Gaussians the accuracy of the variational calculations
increases gradually.

Other angular momenta show a similar behaviour, the convergence is in
general better for lower angular momenta. Thus we can conclude, that
the variational ansatz~(\ref{eq:varansatz}) can describe, though not
quantitatively perfect, the dynamical instabilities of the Bogoliubov
spectrum for angular momenta $m>0$.

\begin{figure}[t]
\centering
  \includegraphics[width=0.7\columnwidth]{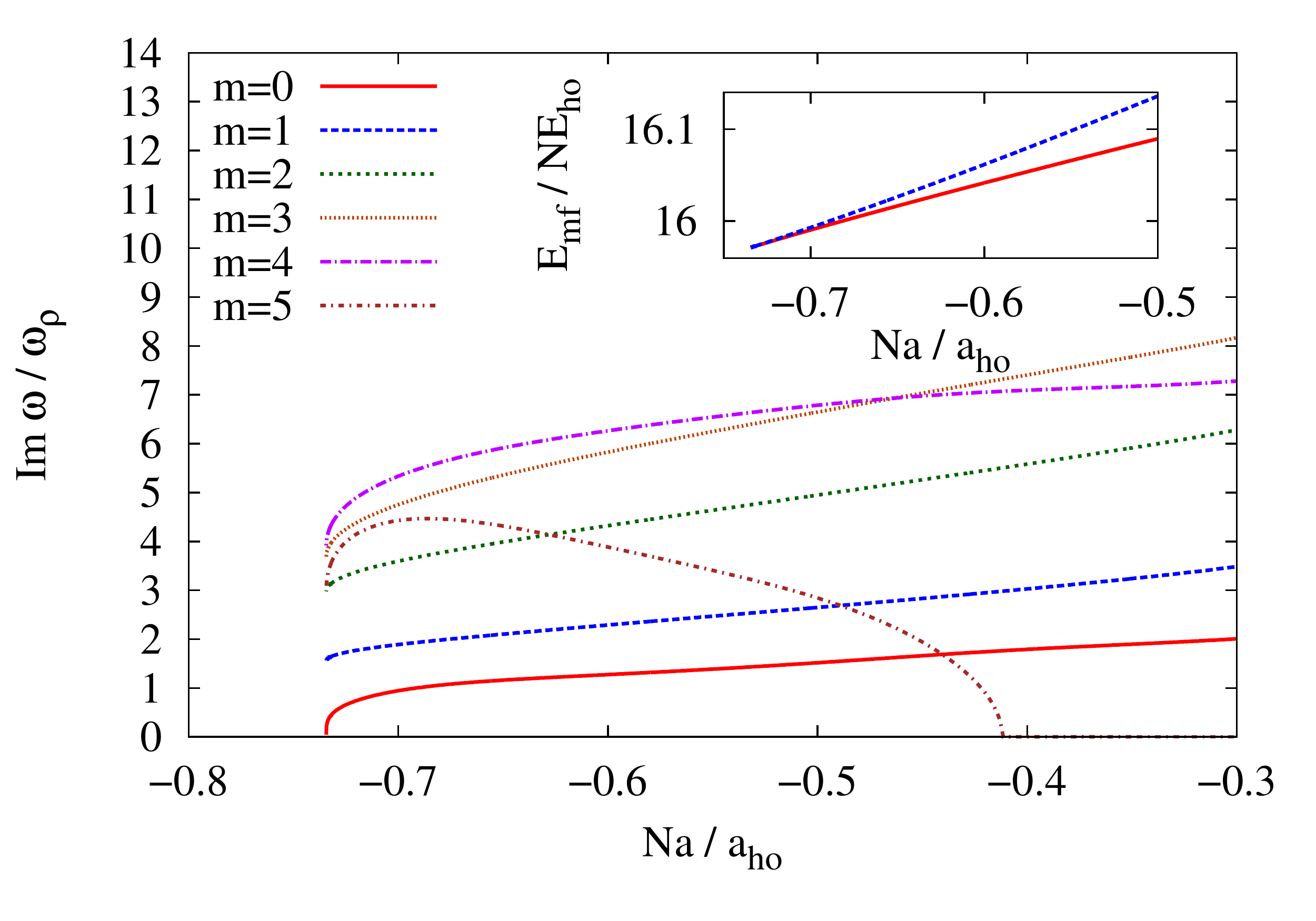}
  \caption{Imaginary frequencies of the excited state with
    $N_\text{G}=5$ Gaussians. At the critical point the excited state
    shares the same instabilities as the ground state with
    additionally an $m=0$ instability, which is typical for the
    collectively excited state. The inset shows the tangent
    bifurcation in the mean field energy $E_\text{mf}/N E_\text{ho}$
    ($N$ being the particle number) for the ground (solid) and excited
    (dashed) state, where $E_\text{ho} = \hbar \omega_\rho$ is the
    energy quantum of the harmonic oscillator.}
  \label{fig:dipolar_max_eigenvals}
\end{figure}

It is known that at the critical scattering length below which no
stationary solution exists anymore an unstable collectively excited
state emerges in a tangent bifurcation \cite{Rau10b} in addition to
the ground state (see the inset in
Fig.~\ref{fig:dipolar_max_eigenvals}). For condensates without
\ac{DDI}, this excited state is unstable with respect to an $m=0$
excitation \cite{Rau10b}. The variational approach offers the
possibility to investigate this state of a dipolar \ac{BEC}, which is
not accessible to the full-numerical \ac{ITE} method, and its collapse
mechanism. Fig.~\ref{fig:dipolar_max_eigenvals} shows the imaginary
frequencies of the excited state obtained with $N_\text{G}=5$
Gaussians. As both states merge at the critical scattering length the
same holds for the eigenfrequencies, and the excited state shares the
instabilities of the ground state at the critical scattering length,
except the additional $m=0$ instability. In contrast to the case with
\ac{SRI} only or with monopolar \ac{LRI} \cite{Kreibich12a} the
excited state of the dipolar \ac{BEC} for the trapping frequencies is
unlikely to collapse with $m=0$ symmetry due to the additional
instabilities for $m>0$, which have a higher imaginary frequency and
thus a larger collapse rate.

The $m=5$ excitation in Fig.~\ref{fig:dipolar_max_eigenvals} changes
its stability with increasing scattering length, in a similar way as
the unstable modes of the ground state (see
Fig.~\ref{fig:dipolar_bogo_gauss_crit}). Note, however, that this does
not mean a global change of the stability due to the other unstable
modes. The excited state stays unstable for all scattering lengths
considered in our calculations.

\subsection{Shape of Bogoliubov excitations}
\label{sec:shape-od-oscill}

\begin{figure}[t]
\centering
  \includegraphics[width=0.7\columnwidth]{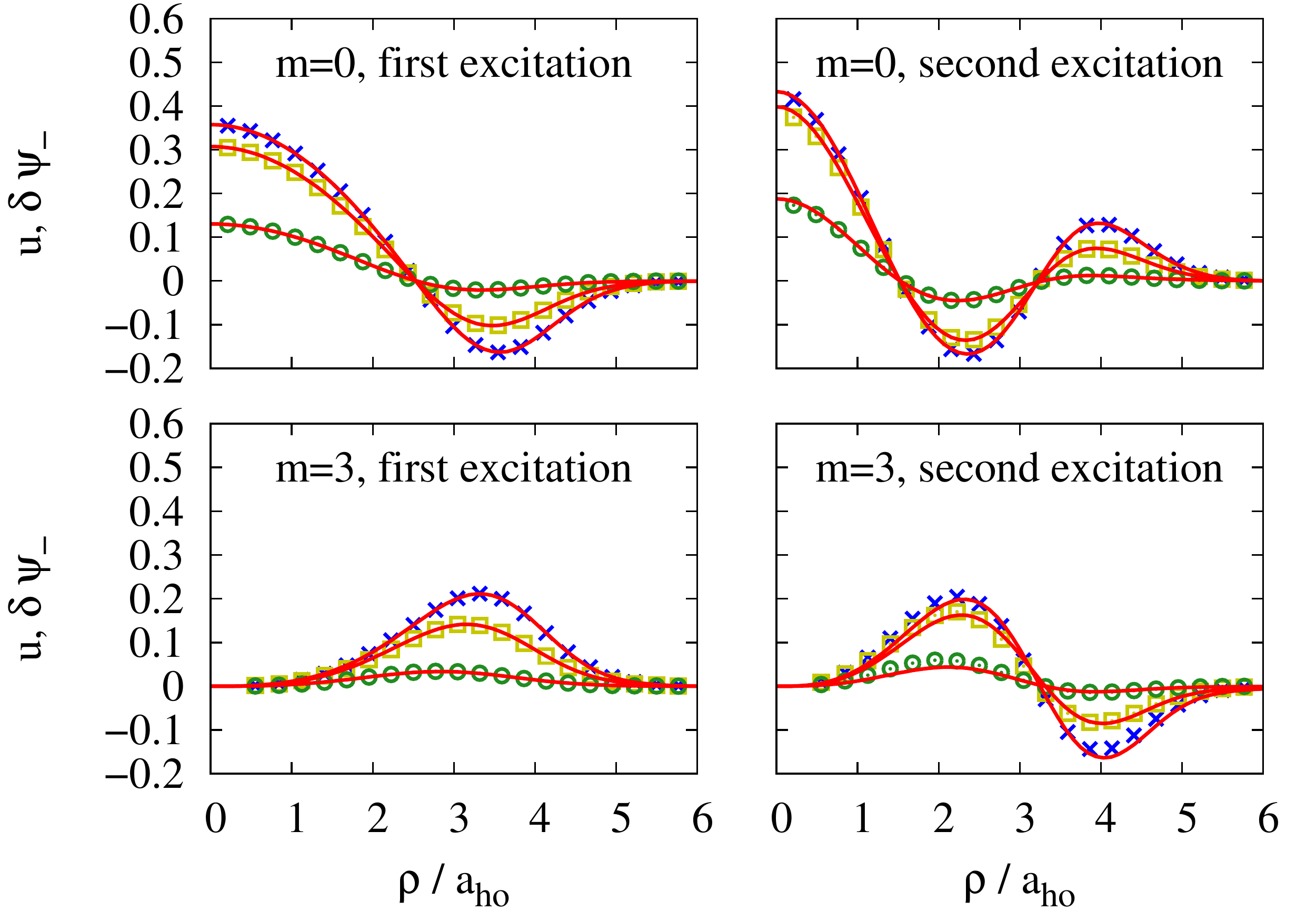}
  \includegraphics[width=0.7\columnwidth]{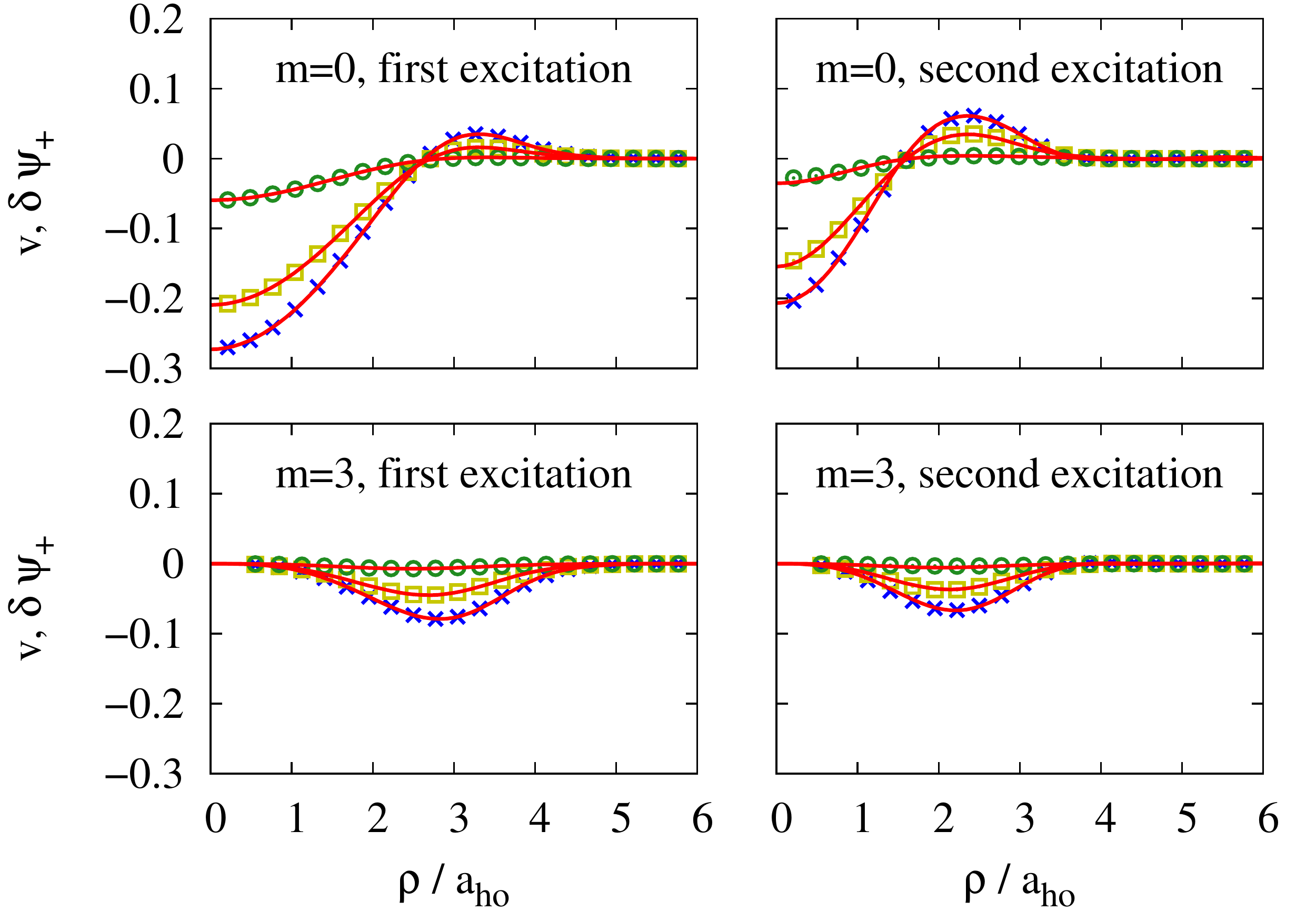}
  \caption{Radial dependence for the functions $u$, $v$ and
    $\delta\psi_-$, $\delta\psi_+$ (red lines), respectively for
    different values of $z$ [$0$ (blue crosses), $0.42 a_\text{ho}$
    (yellow squares), $0.84 a_\text{ho}$ (green circles)], where
    $N_\text{G}=8$ Gaussians have been used (see text for
    normalisation). Shown are the first two excitations for the
    angular momenta $m=0,3$ and a fixed scattering length of
    $Na/a_\text{ho}=6$. All functions are chosen to be real. There is
    a very good agreement, only for the second $m=3$ excitation there
    are small deviations.}
  \label{fig:dipolar_solutions_8gauss_bogo}
\end{figure}

After linking the eigenvalues of the \ac{BDGE} and the Jacobian of the
time-dependent variational approach, we will check if there is also a
direct correspondence between the eigenfunctions and eigenvectors of
these two approaches. In Sec.~\ref{sec:variational-approach} we
discussed how to construct the functions $\delta\psi_-$ and
$\delta\psi_+$ from the eigenvectors and how they are related to the
functions $u$ and $v$. For a better comparison of both methods, we use
the usual normalisation of the Bogoliubov functions $\int \dd^3 r \,
[\abs{u}^2-\abs{v}^2]=1$ \cite{Pitaevskii03a}, and assume the same for
the variational method, $\int \dd^3 r \, [\abs{\delta\psi_-}^2 -
\abs{\delta\psi_+}^2]=1$. We point out that in calculating
$\delta\psi_\pm$ there is no ambiguity, except for a global phase,
which we choose in such a way that the functions are
real. Fig.~\ref{fig:dipolar_solutions_8gauss_bogo} shows the
comparison of both methods for some modes. The first two functions $u$
with $m=0$ have one and two radial nodes, respectively, because the
lowest mode with $m=0$ represents the gauge mode, which is no physical
excitation. For other angular momenta, the $n$-th excitation in $\rho$
direction has $n-1$ radial nodes. The asymptotic behaviour of all
functions in Fig.~\ref{fig:dipolar_solutions_8gauss_bogo} for $\rho
\to 0$ is $\rho^{\abs{m}}$. The nodal structure for the functions $v$
is different. All $v$ for $m=0$ have one node, and for the other
angular momenta zero nodes. For higher excitation numbers and angular
momenta, the amplitude of $v$ becomes smaller and smaller, indicating
that in these limits the coupling between $u$ and $v$ in the \ac{BDGE}
can be neglected, which is in accordance with previous calculations in
the literature \cite{Dalfovo97a,Pitaevskii03a,Ronen06a}.

For the second excitation with $m=3$ we find small deviations, which
is to be expected, since there are also deviations in the
eigenfrequencies (cf.\ Fig.~\ref{fig:dipolar_spectrum}). By comparing
more eigenmodes we can conclude: The better the agreement of the
eigenfunctions, the better is also the accordance of the
frequencies. For higher modes, where there is no quantitative
agreement between the eigenfrequencies, only the asymptotic behaviour
and the nodal structure can be described by the variational ansatz,
even though there is at least a qualitative agreement in those
aspects.

The comparison of the eigenfunctions of the \ac{BDGE} with those
obtained from the eigenvectors of the Jacobian reveals a very good
agreement. This is remarkable, since both functions originate from
quite different approaches of stability. The extended variational
ansatz~(\ref{eq:varansatz}) cannot only quantitatively describe the
frequencies of the lowest modes, but also the shape of the
oscillations, even for angular momenta, which were not accessible by
the variational approach without the extension.

\section{Dynamics}
\label{sec:dynamics}

The extended variational ansatz~(\ref{eq:varansatz}) was proposed to
find a direct relationship between the eigenvalues of the Jacobian and
the eigenfrequencies of the \ac{BDGE}. We have derived the equations
of motion using the \ac{TDVP} to calculate the Jacobian. The equations
of motion~(\ref{eq:tdvpeom}) are not only valid in the vicinity of a
fixed point, but for all variational parameters for which the
integrals in the equations of motion converge. Wilson \etal
\cite{Wilson09a} simulated the dynamical evolution of a collapse for a
dipolar condensate with biconcave structure by reducing the scattering
length below the critical value, which yields a symmetry breaking
collapse (\textit{angular collapse}) with $m=3$ symmetry. Rau \etal
\cite{Rau10b} simulated this scenario within the variational framework
with an ansatz of coupled Gaussians. However, as mentioned in
Sec.~\ref{sec:variational-approach}, this ansatz cannot describe wave
functions that exhibit the $m=3$ symmetry. Motivated by these papers
we will now integrate the equations of motion numerically with
appropriate initial conditions to verify, whether or not the extended
variational ansatz is also applicable in calculating the dynamics of
dipolar condensates and especially to describe the angular collapse
for arbitrary angular momenta, which is caused by the roton
instability (cf.\ Secs.~\ref{sec:full-numer-treatm}
and~\ref{sec:results-dipolar-bec}). To do so we only address the
collapse of ground states with biconcave structure.

\begin{figure}[t]
\centering
  \includegraphics[width=0.23\columnwidth]{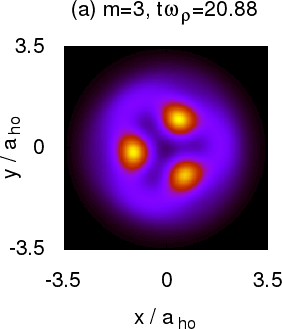}
  \includegraphics[width=0.23\columnwidth]{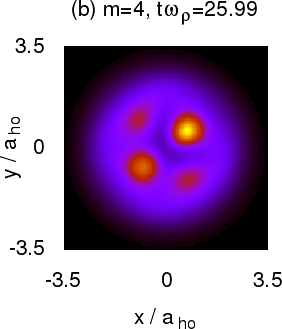}
  \caption{Modulus squared $\abs{\psi}^2$ of the wave function of the
    collapsing dipolar condensate in a cut through the
    $(z=0)$-plane. Bright (yellow) marks regions of high particle
    density, while dark (black) stands for low densities. For the
    calculations $N_\text{G}=4$ Gaussians were used.}
  \label{fig:dipolar_collapse_m}
\end{figure}

As a first example, we consider the system with $N_\text{G}=4$
Gaussians and calculate the dynamics of an induced collapse by
lowering the scattering length below the critical value. As an initial
state for $t=0$, we take the stable ground state for a scattering
length of $Na/a_\text{ho}=6$ and add a small perturbation for a
specific angular momentum by changing the variational parameter
$d_{m,p=0}^{k=1}$ to a small value larger than zero, e.g.\
$d_{m,p=0}^{k=1}=2 \times 10^{-5}$, which is just a small perturbation
of the ground state. We let the condensate evolve until $t\omega_\rho
= 0.72$, after which the scattering length is linearly ramped down
until $t\omega_\rho = 10.8$ to a value of $Na/a_\text{ho}=-0.6$, which
is below the critical scattering length. The condensate then evolves
at a fixed scattering length, until it collapses to a point, from
where on no further integration is possible.  We calculated this
scenario for an initial $m=3$ and $m=4$ excitation, which cannot be
described by the variational ansatz without the extension and thus is
an advance as compared to previous calculations of a collapsing
condensate within the variational framework \cite{Rau10b}.

Fig.~\ref{fig:dipolar_collapse_m} shows the resulting density of the
collapsing condensate for both cases. For an initial $m=3$ excitation,
the collapsing cloud exhibits the clear shape with an $m=3$ symmetry,
the angular dependency is proportional to $\sin(3\phi)$. This result
shows that the extended variational ansatz is not only able to
describe the instability in the vicinity of the ground state, but also
the nonlinear dynamics given by the equations of
motion~(\ref{eq:tdvpeom}). If the initial state is perturbed by an
$m=4$ excitation, the resulting density distribution shows a
superposition of several angular momenta. There are altogether four
density peaks, of which two are significantly larger than the other
two. The appearance of four peaks corresponds to an $m=4$ symmetry,
but the distribution of the particle density clearly belongs to
$m=2$. Of the two large peaks, one is slightly higher than the other
one which indicates an $m=1$ symmetry. The calculation shows that the
different angular momenta do not evolve freely, but instead are
coupled and influence each other. This is due to the nonlinearity of
the \ac{GPE}~(\ref{eq:tgpe}) which carries over to the equations of
motion for the variational parameters.

There are many excitations one can add to the ground state as a
perturbation. Above, we added specific eigenmodes to investigate the
behaviour of the extended variational ansatz when considering the
collapse dynamics. We now want to utilise a more physical motivated
initial state. In Ref.~\cite{Wilson09a} the authors added modes with
weights determined by the Bose-Einstein distribution, which leads to
\begin{align}
  \label{eq:perturbtemp}
  \psi_0(\rho,z) + \sum\limits_{n,m} \sqrt{\frac{n_{n,m}}{N}} \eto{2
    \pi \ii \alpha_{n,m}} \left[ u_{n,m}(\rho,z) \eto{\ii m \phi} +
    v_{n,m}^*(\rho,z) \eto{-\ii m \phi} \right],
\end{align}
where $\psi_0$ is the ground state. The weight of each excitation,
where $m$ is the angular momentum and $n$ summarises the other quantum
numbers, is given by
\begin{align}
  n_{n,m} = \left( \eto{\hbar\omega_{n,m}/k_\text{B} T}-1
  \right)^{-1}.
\end{align}
Here, $\omega_{n,m}$ is the energy of each excitation, $k_\text{B}$
the Boltzmann constant, and $T$ the temperature. The quantities
$\alpha_{n,m}$ in Eq.~(\ref{eq:perturbtemp}) determine the initial
phase of each excitation and are random numbers between $0$ and
$1$. The overall factor $1/\sqrt{N}$, with $N$ the particle number, is
necessary, since the wave function $\psi$ is normalised to unity
instead of $N$.

\begin{figure}[t]
\centering
  \includegraphics[width=0.23\textwidth]{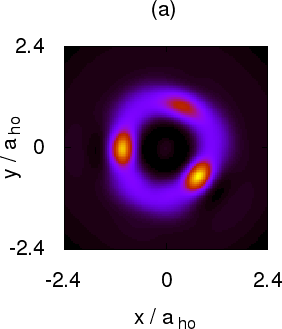}
  \includegraphics[width=0.23\textwidth]{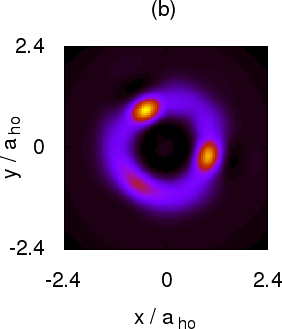}
  \includegraphics[width=0.23\textwidth]{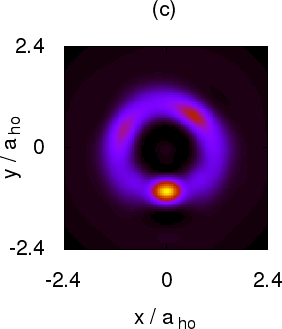}
  \includegraphics[width=0.23\textwidth]{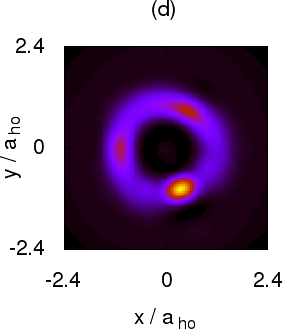}
  \caption{Particle densities of the collapsing cloud at $t
    \omega_\rho = 12.384$ with high (low) density represented by
    bright (dark) regions. All calculations were done with
    $N_\text{G}=5$ Gaussians. The different random phases used to
    construct the initial state (see text) are the only differences in
    the four calculations. There are three density peaks each, but the
    overall rotation and the density distribution along the peaks are
    different.}
  \label{fig:dipolar_collapse_5g_m6}
\end{figure}

We now can make use of the relationship between the eigenvectors of
the Jacobian and the eigenfunctions of the \ac{BDGE}
(Sec.~\ref{sec:shape-od-oscill}) to prepare an initial state similar
to that given by Eq.~(\ref{eq:perturbtemp}) within the variational
framework, where instead of the functions, the eigenvectors of the
variational parameters have to be added to the ground state
vector. For the following calculations we simulate a condensate
consisting of $N=10000$ \ce{^{52}Cr} atoms at a temperature of
$T=\unit{100}{\nano\kelvin}$. We start with an initial wave function
according to Eq.~(\ref{eq:perturbtemp}) and a scattering length of
$Na/a_\text{ho} = 6$, which is ramped down to a value of
$Na/a_\text{ho} = -3$ over a time of $\Delta t \omega_\rho = 14.68$.

We performed several calculations with this scenario, for which the
only differences are the initial phases $\alpha_{n,m}$. As the
scattering length decreases the particle density establishes its
biconcave structure, which is a typical feature of dipolar \acp{BEC}
\cite{Wilson09a,Rau10c,Ronen07a,Dutta07a}. If the scattering length
further decreases, the critical value is reached and the atomic cloud
starts to collapse to the region with the highest density, which is on
a ring around the centre. Due to the initial excitations with angular
momentum $m>0$, which are now unstable, the symmetry is broken and
several density peaks emerge. Fig.~\ref{fig:dipolar_collapse_5g_m6}
shows the density distribution of four different calculations shortly
before collapse. Each condensate has three peaks located almost
equidistantly on a ring. But the distribution along the ring differs:
In Figs.~\ref{fig:dipolar_collapse_5g_m6} (a)-(b) there are two large
peaks and a smaller one, while in
Figs.~\ref{fig:dipolar_collapse_5g_m6} (c)-(d) it is the other way
round. Additionally the initial phases determine the overall rotation
of the collapsed condensate in the $(z=0)$-plane. We can conclude that
both, the $m=2$ and $m=3$ mode, are responsible for the shape of the
collapsed cloud, which corresponds to the numerical solution of the
\ac{BDGE} (see Fig.~\ref{fig:dipolar_bogo_crit}), in which both modes
turn unstable almost simultaneously.

A question, which naturally arises, is how the number of Gaussians
influences the dynamics calculation. We performed similar calculations
with $N_\text{G}=4$ and $N_\text{G}=6$ Gaussians. With $4$ Gaussians
we find a different behaviour in that a collapse with $m=2$ symmetry
is favoured. However, with $6$ Gaussians almost the same results as in
the case with $5$ Gaussians are obtained. The only difference is the
time taken before the condensate collapses, which is shorter with $6$
Gaussians. This can be explained by the fact that the critical
scattering length is shifted to a higher value and thus the critical
point is reached earlier during the simulation. The eigenfrequencies
with our Gaussian ansatz do not fully agree with the full-numerical
results (cf.\ Sec.~\ref{sec:results-dipolar-bec}), however, we expect
the variational results to not being completely converged with $5$ or
$6$ Gaussians.

\section{Conclusion and outlook}
\label{sec:conclusion-outlook}

We presented an extended variational ansatz, which is specific to
cylindrically symmetric systems and which can describe arbitrary
angular momenta, and derived the equations of motion for the
variational parameters using the \ac{TDVP}. The eigenfrequencies of
two different systems (without and with \ac{DDI}) were calculated
using the variational ansatz and compared with the direct solutions of
the \ac{BDGE}. We found a good agreement of the lowest-lying modes and
could confirm the direct relationship of both methods. This extended
our investigations of spherically symmetric systems \cite{Kreibich12a}
and we could show that this agreement is of generic type, independent
of the symmetry and the interactions.

The roton-instability of a pancake-shaped dipolar \ac{BEC} was
investigated using the extended ansatz. We could find instabilities
for arbitrary values of the angular momenta, which is a great progress
in comparison with the variational ansatz with coupled Gaussians
without the extension. With more and more Gaussians we obtained
gradually more accurate results within the variational
framework. Thus, our variational method is a valid approximation to
grid calculations.

We further constructed functions of the eigenvectors of the Jacobian
and found that these functions agree very well with the solutions of
the \ac{BDGE}, as long as the corresponding eigenvalues agree. There
is not only a direct link between the eigenfrequencies, but also
between the eigenvectors and the eigenfunctions.

After considering the dynamics in the linear regime we directly
integrated the nonlinear equations of motion and showed that our
variational ansatz, assuming that the initial conditions and the
scenario of the time-dependent scattering length are appropriately
chosen, is capable of describing the angular collapse with arbitrary
angular momenta. We used the relationship between the eigenvectors of
the Jacobian and the eigenfunctions of the \ac{BDGE} to construct an
initial state which corresponds to a condensate in thermal equilibrium
within the mean field approximation, and simulated a realistic
experiment of the angular collapse. We found that the random initial
phase distribution strongly influences the shape of the collapsed
atomic cloud, which makes the resulting shape of an experiment
irreproducible.

Altogether it was shown that the extended variational ansatz is a
significant progress compared to coupled Gaussians, and as yet
unaccessible results could be obtained. Our ansatz is a valid
alternative to calculate eigenfrequencies and shapes of elementary
excitations, if the lowest eigenmodes of stable condensates are
considered. The roton-instabilities can be obtained, but we could not
find a good quantitative agreement, leaving our ansatz an
approximation.

In future applications of the extended variational ansatz it would be
possible to include the particle loss in the dynamics calculation, as
the losses are necessary to accurately simulate a collapse scenario of
a dipolar \ac{BEC} \cite{Wilson09a, Lahaye08a}. Furthermore, our
ansatz is applicable to dipolar \acp{BEC} in a one-dimensional optical
lattice \cite{Koeberle09b, Junginger10a, Wilson11a} to calculate the
ground state, stability, excitations and the dynamics. The direct
relationship between the eigenvectors of the Jacobian and the
eigenfunctions of the \ac{BDGE} allows several applications of the
variational approach, e.\,g.\ to describe a \ac{BEC} at finite
temperature, where a self-consistent solution of the \ac{GPE} and
\ac{BDGE} is necessary \cite{Proukakis08a}.

\ack

This work was supported by Deutsche Forschungsgemeinschaft.

\appendix

\section{Calculation of the integrals for the variational ansatz}
\label{sec:calc-integr}

This appendix is dedicated to the integrals necessary for the matrix
$\mathbf{K}$ and the vector $\vec{h}$, which are defined by
\begin{subequations}
  \begin{align}
    K_{ij} &=
    \qmprod{\firstpderiv{\psi}{z_i}}{\firstpderiv{\psi}{z_j}} = \int
    \dd^3 r \left( \firstpderiv{\psi}{z_i} (\vec{r}) \right)^*
    \firstpderiv{\psi}{z_j} (\vec{r}), \\
    h_i &= \qmprod{\firstpderiv{\psi}{z_i}}{\hat{H} \psi} = \int \dd^3
    r \left( \firstpderiv{\psi}{z_i} (\vec{r}) \right)^* \hat{H}
    \psi(\vec{r}),
  \end{align}
\end{subequations}
to set up the equations of motion~(\ref{eq:tdvpeom}) of the
variational ansatz~(\ref{eq:varansatz}) via the time-dependent
variational principle. We write the variational ansatz in the form
\begin{align}
  \psi = \sum\limits_{k=1}^{N_\text{G}} \sum\limits_m
  \sum\limits_{p=0,1} d_{m,p}^k \rho^{\abs{m}} z^p \eto{\ii m \phi}
  \eto{-A_\rho^k \rho^2 - A_z^k z^2 - p_z^k z -\gamma^k},
\end{align}
where $d_{0,0}^k \equiv 1$ and $d_{0,1}^k \equiv 0$ are no variational
parameters, but constants. We further introduce the abbreviations
\begin{align}
  \qmprod{z_i}{z_j} \equiv
  \qmprod{\firstpderiv{\psi}{z_i}}{\firstpderiv{\psi}{z_j}}, &&
  \qmprod{z_i}{X \psi} \equiv \qmprod{\firstpderiv{\psi}{z_i}}{X\psi}
\end{align}
for the matrix elements, where $X$ is an operator of the mean field
Hamiltonian $H = T + V_\text{ext} + \Phi_\text{s} + \Phi_\text{d}$
with the kinetic part, the external potential and the scattering and
dipolar mean field potentials.

We do not give a detailed way on how to calculate all necessary
integrals, nor do we present all results. We rather want to sketch the
calculations and give results of some examples. For all details we
refer to the master thesis of one of the authors \cite{Kreibich11a}.

\subsection{Integrals of the K-matrix}
\label{sec:integrals-k-matrix}

Before giving the results, we define the abbreviations
\begin{align}
  A_\rho^{kl} \equiv A_\rho^k + (A_\rho^l)^*, && A_z^{kl} \equiv A_z^k
  + (A_z^l)^*, && p_z^{kl} \equiv p_z^k + (p_z^l)^*, && \gamma^{kl}
  \equiv \gamma^k + (\gamma^l)^*
\end{align}
and the basic integrals
\begin{align}
  \label{eq:intansatz}
  I^{kl}(m,n) &\equiv \int\limits_0^\infty \dd \rho \, \rho^{2m-1}
  \eto{-A_\rho^{kl} \rho^2} \int\limits_{-\infty}^\infty \dd z \, z^n
  \eto{-A_z^{kl} z^2 - p_z^{kl} z} \eto{-\gamma^{kl}} =
  \eto{-\gamma^{kl}} \frac{\Gamma(m)}{2 (A_\rho^{kl})^m} I_z^{kl}(n).
\end{align}
The $\rho$ integral can be identified with the gamma function by a
simple substitution. The $z$ integral is a Gaussian integral for
$n=0$, higher orders can be obtained by the recursion formula
\begin{align}
  I_z^{kl}(n+1) = -\firstpderiv{}{p_z^{kl}} I_z^{kl}(n).
\end{align}
The integrals of the matrix $\mathbf{K}$ then yield
\begin{subequations}
  \begin{align}
    \qmprod{d_{m_2,p_2}^l}{d_{m_1,p_1}^k} &= 2\pi \delta_{m_1 m_2}
    I^{kl}(\abs{m_1}+1,p_1+p_2), \\
    \qmprod{d_{m_2,p_2}^l}{A_\rho^k} &= -2\pi \sum\limits_{p_1}
    I^{kl}(\abs{m_2}+2,p_1+p_2), \\
    \qmprod{d_{m_2,p_2}^l}{A_z^k} &= -2\pi \sum\limits_{p_1}
    I^{kl}(\abs{m_2}+1,p_1+p_2+2), \\
    \qmprod{d_{m_2,p_2}^l}{p_z^k} &= -2\pi \sum\limits_{p_1}
    I^{kl}(\abs{m_2}+1,p_1+p_2+1), \\
    \qmprod{d_{m_2,p_2}^l}{\gamma^k} &= -2\pi \sum\limits_{p_1}
    I^{kl}(\abs{m_2}+1,p_1+p_2).
  \end{align}
\end{subequations}
The other integrals can similarly be expressed by
Eq.~(\ref{eq:intansatz}).

\subsection{Integrals of the kinetic term}
\label{sec:integr-kinet-term}

For the integrals of the kinetic term the action of the Laplacian on
the variational ansatz is necessary. This is most easily evaluated in
cylindrical coordinates. The integrals can then be expressed by
Eq.~(\ref{eq:intansatz}). For the $d$ parameters, e.\,g., one obtains
\begin{multline}
  \qmprod{d_{m_2,p_2}^l}{T \psi} / \hbar \omega_\rho = \pi
  \sum\limits_k \sum\limits_{p_1} d_{m_2,p_1}^k \Bigl\{ \left[ 4
    (\abs{m_2}+1) A_\rho^k + 2(2p_1+1) A_z^k - (p_z^k)^2
  \right] \\
  \times I^{kl}(\abs{m_2}+1,p_1+p_2) +\left[ 2 p_1 p_z^k \right]
  I^{kl}(\abs{m_2}+2,p_2) -\left[ 4 p_z^k A_z^k
  \right] I^{kl}(\abs{m_2}+2,p_1+p_2+1) \\
  -\left[ 4 (A_z^k)^2 \right] I^{kl}(\abs{m_2}+2,p_1+p_2+2) -\left[ 4
    (A_\rho^k)^2 \right] I^{kl}(\abs{m_2}+3,p_1+p_2) \Bigr\}.
\end{multline}

\subsection{Integrals of the trapping potential}
\label{sec:integr-trapp-potent}

The trapping potential is of the form $V_\text{ext} / = \frac{M}{2}
\omega_\rho^2 (\rho^2 + \lambda^2 z^2)$. Thus, the corresponding
integrals are easy to evaluate. As an example we obtain
\begin{multline}
  \qmprod{d_{m_2,p_2}^l}{V_\text{ext} \psi} / \hbar \omega_\rho=
  \pi \sum\limits_{k} \sum\limits_{p_1} d_{m_2,p_1}^k \\
  \times \Bigl( I^{kl}(\abs{m_2}+2,p_1+p_2) + \lambda^2
  I^{kl}(\abs{m_2}+1,p_1+p_2+2) \Bigr).
\end{multline}

\subsection{Integrals of the scattering interaction}
\label{sec:integr-scatt-inter}

The integrals for the interaction terms require some more
attention. First an expression for the modulus squared wave function
$\abs{\psi}^2$ is needed. This enters the integrals for the
interaction potentials. The scattering interaction is easier to
calculate, since this kind of interaction is local, in contrast to the
\ac{DDI}.

We further need new abbreviations, which depend on four indices and
which are defined by
\begin{align}
  A_\rho^{ijkl} \equiv A_\rho^{ij} + A_\rho^{kl}, && A_z^{ijkl} \equiv
  A_z^{ij} + A_z^{kl}, && p_z^{ijkl} \equiv p_z^{ij} + p_z^{kl}, &&
  \gamma^{ijkl} \equiv \gamma^{ij} + \gamma^{kl}.
\end{align}
The new basic integral, $I^{ijkl}(m,n)$, can be obtained from the
former one, Eq.~(\ref{eq:intansatz}), by substituting all indices $kl$
with $ijkl$. The scattering integral then reads
\begin{multline}
  \qmprod{d_{m_2,p_2}^l}{\Phi_\text{s} \psi} / \hbar \omega_\rho = 8
  \pi^2 \frac{N a}{a_\text{ho}} \sum\limits_{i,j,k}
  \sum\limits_{m_1,m_3,m_4} \sum\limits_{p_1,p_3,p_4}
  (d_{m_4,p_4}^j)^* d_{m_3,p_3}^i d_{m_1,p_1}^k
  \delta_{m_1+m_3,m_2+m_4} \\
  \times I^{ijkl} \left(
    \frac{\abs{m_1}+\abs{m_2}+\abs{m_3}+\abs{m_4}+2}{2},
    p_1+p_2+p_3+p_4 \right).
\end{multline}

\subsection{Integrals of the dipolar interaction}
\label{sec:integr-dipol-inter}

In order to calculate the integrals of the dipolar interaction, we
first need the expression for the mean field potential
$\Phi_\text{d}$. This potential can be rewritten using Fourier
transforms and the convolution theorem, which yields \cite{Goral02a}
\begin{align}
  \Phi_\text{d}(\vec{r}) =
  \mathcal{F}^{-1}[\tilde{V}_\text{d}(\vec{k}) \cdot
  \tilde{n}(\vec{k})],
\end{align}
where the functions with tilde designate the Fourier transforms of the
dipole potential and the particle density. The transform of the dipole
potential is well-known to be \cite{Goral02a}
\begin{align}
  \tilde{V}_\text{d}(\vec{k}) / \hbar \omega_\rho = \frac{4\pi}{3}
  \left( 3 \frac{k_z^2}{k^2} - 1 \right) D/N.
\end{align}
Our next task is to calculate $\tilde{n}(\vec{k})$. Writing the square
of the absolute value of the wave function with the given variational
ansatz, one notices that the angular dependency is of the form
$\exp(\ii m \phi)$ with some $m \in \mathbb{Z}$. In \cite{Ronen06a} it
was noted that the three-dimensional Fourier transform of such a
function can be expressed as a one-dimensional Hankel transform for
the $\rho$ direction and a remaining one-dimensional Fourier transform
in the $z$ direction. The Hankel transform brings in the Bessel
function, and the integral can be worked out by Taylor expanding the
Bessel function. The Fourier transform is a Gaussians integral with
linear terms and variable powers of $z$, for which an analytical
expression can be found in established tables of integrals
\cite[Eq.~3.462.2]{Gradshteyn94a}.

In the original integral, $\qmprod{d}{\Phi_\text{d} \psi}$, the
integral over $\vec{r}$ is similar to the Fourier transform of
$\abs{\psi}^2$ and can easily be expressed in the same way, which
yields the intermediate result
\begin{multline}
  \qmprod{d_{m_2,p_2}^l}{\Phi_\text{d} \psi} = \frac{1}{2\pi}
  \sum\limits_{i,j,k} \sum\limits_{m_1,m_3,m_4}
  \sum\limits_{p_1,p_3,p_4} \eto{-\gamma^{ijkl}} (d_{m_4,p_4}^j)^*
  d_{m_3,p_3}^i d_{m_1,p_1}^k \sigma_{m_1-m_2} \sigma_{m_3-m_4} \\
  \times
  \frac{\ii^{m_1-m_2-m_3+m_4}}{2^{\abs{m_1-m_2}+\abs{m_3-m_4}+2}}
  (A_\rho^{kl})^{-\abs{m_1-m_2}-\mu_{m_1,m_2}-1}
  (A_\rho^{ij})^{-\abs{m_3-m_4}-\mu_{m_3,m_4}-1} \\
  \times \sum\limits_{\alpha=0}^{\mu_{m_1,m_2}}
  \sum\limits_{\beta=0}^{\mu_{m_3,m_4}}
  \begin{pmatrix} \mu_{m_1,m_2} \\ \alpha \end{pmatrix}
  \begin{pmatrix} \mu_{m_3,m_4} \\ \beta \end{pmatrix} (4
  A_\rho^{kl})^\alpha (4 A_\rho^{ij})^{\beta}
  (-1)^{\mu_{m_1,m_2}+\mu_{m_3,m_4}} \\
  \times (-\abs{m_1-m_2}-\mu_{m_1,m_2})_{\mu_{m_1,m_2}-\alpha} \,
  (-\abs{m_3-m_4}-\mu_{m_3,m_4})_{\mu_{m_3,m_4}-\beta} \, \times I_k,
\end{multline}
where we defined
\begin{align}
  \sigma_m \equiv (\sign m)^m, && \mu_{m_1,m_2} \equiv
  \begin{cases}
    0 & \text{for $\sign m_1 \neq \sign m_2$}, \\
    \min(\abs{m_1},\abs{m_2}) & \text{otherwise},
  \end{cases}
\end{align}
and used the Pochhammer symbol $(a)_n$, which is defined by
\begin{align}
  (a)_n \equiv a (a-1) (a-2) \cdots (a-n+1) =
  \frac{\Gamma(a+1)}{\Gamma(a-n+1)}.
\end{align}
The remaining three dimensional integral $I_k$ reads
\begin{multline}
  I_k = \int \dd^3 k \, \eto{\ii(m_1+m_3-m_2-m_3)k_\phi}
  \eto{-k_\rho^2 / 4 A_\rho^{ij}} \eto{-k_\rho^2 / 4 A_\rho^{kl}} \\
  \times k_\rho^{2(\alpha+\beta)+\abs{m_1-m_2}+\abs{m_3-m_4}}
  \tilde{I}_{-z}^{kl}(p_1+p_2) \tilde{I}_z^{ij}(p_3+p_4)
  \tilde{V}_\text{d}(\vec{k}),
\end{multline}
with the $z$ integrals
\begin{subequations}
  \begin{align}
    \tilde{I}_z^{ij}(p) &\equiv \int\limits_{-\infty}^\infty \dd z \,
    z^p \eto{-A_z^{ij} z^2 - (p_z^{ij}+\ii k_z) z}, \\
    \tilde{I}_{-z}^{kl}(p) &\equiv \int\limits_{-\infty}^\infty \dd z
    \, z^p \eto{-A_z^{kl} z^2 - (p_z^{kl}-\ii k_z) z},
  \end{align}
\end{subequations}
which are evaluated in Ref.~\cite[Eq.~3.462.2]{Gradshteyn94a}. The
Fourier transform of the dipole potential consists of two parts,
$V_\text{d} = (4\pi k_z^2 / k^2 - 4\pi/3) \hbar \omega_\rho D/N \equiv
V_\text{d}^1 + V_\text{d}^2$. The integral of $I_k$ containing
$V_\text{d}^2$ is equivalent to the integrals of the scattering
interaction\footnote{The Fourier transform of $V_\text{s}=4\pi Na /
  a_\text{ho} \delta(\vec{r})$ is $4\pi Na / a_\text{ho}$, and thus
  proportional to $V_\text{d}^2$. This means that the whole integral
  of this part is equivalent to the integrals of the scattering
  interaction.}, and requires no further attention. The part of $I_k$
with $V_\text{d}^1$, which we denote by $I_k^1$, can be rewritten to
\begin{align}
  I_k^1 / \hbar \omega_\rho &= 4\pi^2 (D/N) \delta_{m_1+m_3,m_2+m_4}
  n! (\alpha_\rho^{ijkl})^n \nonumber \\
  &\times\int\limits_1^\infty \dd t \, t^{-(n+1)}
  \int\limits_{-\infty}^\infty \dd k_z \, k_z^2 \eto{-(t-1)
    \alpha_\rho^{ijkl} k_z^2} \tilde{I}_{-z}^{kl}(p_1+p_2)
  \tilde{I}_z^{ij}(p_3+p_4).
\end{align}
The $k_z$ integral is of Gaussian type, and similar to
$\tilde{I}_{-z}^{kl}$ and $\tilde{I}_{z}^{ij}$. The remaining $t$
integral can be expressed by the hypergeometric Function $_2 F_1$
\cite{Abramowitz72a}, assuming all parameters $p_z^k$ are zero. In the
general case, the $t$ integral can be efficiently evaluated using a
Taylor series.

\section{Structure of the Jacobian}
\label{sec:structure-jacobian}

The solutions of the \ac{BDGE}~(\ref{eq:bdg}) can be classified by the
quantum numbers $m$ and $\pi_z$, which express the symmetries of the
system. Applying the \ac{TDVP} and linearizing the equations of motion
yields an alternative description of linear oscillations and
stability, in which the Jacobian $\mathbf{J}$ plays a central
role. The Jacobian, similar to the \ac{BDGE}, also express the
symmetries of the system and the quantum numbers, in that it assumes a
block structure, which can in general be written as
\begin{align}
  \mathbf{J} =
  \begin{pmatrix}
    \{ \mathbf{J}_{\pm m}^{\pi_z} \}
  \end{pmatrix} =
  \begin{pmatrix}
    \mathbf{J}_0^{+1} \\
    & \mathbf{J}_0^{-1} \\
    & & \mathbf{J}_{\pm 1}^{+1} \\
    & & & \mathbf{J}_{\pm 1}^{-1} \\
    & & & & \ddots
  \end{pmatrix},
\end{align}
where $\mathbf{J}_{\pm m}^{\pi_z}$ is the submatrix belonging to the
angular momenta $\pm m$ and the parity $\pi_z$. This block structure
allows one to set up and diagonalise the Jacobian for each $m$ and
$\pi_z$ individually, analogously to solving the \ac{BDGE}. The only
difference is that in the Jacobian the angular momenta $m$ and $-m$
are coupled, which is due to the coupling of those angular momenta in
the integrals necessary for the equations of motion
(cf.~\ref{sec:calc-integr}).

A simple example, which shows that the coupling may not be neglected,
is the submatrix of the $m=\pm 1$, $\pi_z=1$ Jacobian obtained with
$N_\text{G}=1$ Gaussian, for a dipolar \ac{BEC} with $\lambda=7$ and
$D=30$
\begin{align}
  \label{eq:matex}
  \mathbf{J}_{\pm 1}^{+1} =
  \begin{pmatrix}
    0 & -2.5936 & 0 & -2.3931 \\
    2.5936 & 0 & -2.3931 & 0 \\
    0 & -2.3931 & 0 & -2.5936 \\
    -2.3931 & 0 & 2.5936 & 0
  \end{pmatrix} \omega_\rho,
\end{align}
where the variational parameters are ordered as $\tilde{\vec{z}} =
(\Real d_{1,0}^0, \Imag d_{1,0}^0, \Real d_{-1,0}^0, \Imag
d_{-1,0}^0)$ for a scattering length of $Na/a_\text{ho} = 6$. The
eigenvalues of the matrix~(\ref{eq:matex}) are $(+\ii \omega_\rho,
+\ii \omega_\rho, -\ii \omega_\rho, -\ii \omega_\rho)$. The
eigenvectors of the eigenvalue $+\ii \omega_\rho$ can then be combined
to an eigenvector belonging to $m=1$ and another eigenvector for
$m=-1$, the same holds for the eigenvalue $-\ii \omega_\rho$. In
general, for arbitrary number of Gaussians and angular momenta, by
taking appropriate linear combinations of the eigenvectors,
eigenvectors belonging to a specific angular momentum $+m$ or $-m$ can
be obtained.

\section*{References}

\bibliographystyle{unsrt}

\end{document}